\documentclass[12pt]{article}

\usepackage[dvips]{graphicx}

\newcommand{\lsim}{\raisebox{-4pt}{$\,\stackrel{\textstyle
                                                         <}{\sim}\,$}}
\newcommand{\gsim}{\raisebox{-4pt}{$\,\stackrel{\textstyle
                                                         >}{\sim}\,$}}
\newcommand{\nn}{\nonumber}
\newcommand{\be}{\begin{equation}}
\newcommand{\ee}{\end{equation}}
\newcommand{\ba}{\begin{eqnarray}}
\newcommand{\ea}{\end{eqnarray}}
\newcommand{\req}[1]{(\ref{#1})}
\def\={\,=\,}
\newcommand{\ci}[1]{\cite{#1}}

\def\mev{~{\rm MeV}}
\def\gev{~{\rm GeV}}

\def\ale{\alpha_{\rm elm}}

\def\eps{\epsilon}

\def\xbj{x_{\rm Bj}}

\newcommand{\tw}{\textwidth}

\def\vb0{{\bf b}_0}

\def\xbj{x_{\rm Bj}}

\def\={\,=\,}

\begin{document} 
\thispagestyle{empty}
\begin{flushright}
WU B 14-03 \\
July, 4  2014\\[20mm]
\end{flushright}

\begin{center}
{\Large\bf The  pion pole in hard exclusive vector-meson leptoproduction}\\
\vskip 10mm

S.V.\ Goloskokov
\footnote{Email:  goloskkv@theor.jinr.ru}
\\[1em]
{\small {\it Bogoliubov Laboratory of Theoretical Physics, Joint Institute
for Nuclear Research,\\ Dubna 141980, Moscow region, Russia}}\\

\vskip 5mm
P.\ Kroll \footnote{Email:  kroll@physik.uni-wuppertal.de}
\\[1em]
{\small {\it Fachbereich Physik, Universit\"at Wuppertal, D-42097 Wuppertal,
Germany}}\\
and
{\small {\it Institut f\"ur Theoretische Physik, Universit\"at
    Regensburg, \\D-93040 Regensburg, Germany}}\\

\end{center}
\vskip 5mm 
\begin{abstract}
\noindent Exploiting a set of generalized parton distributions (GPDs) derived from 
analyses of hard exclusive leptoproduction of $\rho^0$, $\phi$ and $\pi^+$ mesons, we 
investigate the $\omega$ spin density matrix elements (SDMEs) recently measured by 
the HERMES collaboration. It turns out from our study that the pion pole is an 
important contribution to $\omega$ production. It will be treated as a one-particle 
exchange since its evaluation from the GPD $\widetilde E$ considerably underestimates
its contribution. As an intermediate step of our analysis we extract the $\pi\omega$ 
transition form factor for photon virtualities less than $4\,\gev^2$. From our
approach we achieve results for the $\omega$ SDMEs in good agreement with the HERMES 
data. The role of the pion pole in exclusive $\rho^0$ and $\phi$ leptoproduction is 
discussed too.
\end{abstract}   

\section{Introductory remarks}
It is known for a long time that pion exchange plays an important role in photo-
and leptoproduction of $\omega$ mesons \ci{fraas} (see for instance the review 
\ci{bauer78}). Pion exchange also contributes to other reactions, as for instance, 
to exclusive $\omega$ production in proton-proton collisions at high energies 
\ci{cisek14}. The residue of the pion pole in $\omega$ leptoproduction includes the 
$\gamma^* \pi\omega$ vertex function. An analysis of the pertinent processes therefore 
allows for an extraction of information on this vertex. The recent HERMES measurement 
\ci{hermes-omega} of the SDMEs for electroproduced $\omega$ meson at fairly large 
values of the photon virtuality, $Q^2$, small Bjorken-$x$, $\xbj$, and small invariant 
momentum transfer, $t$, offers a unique possibility to learn about the  
$\gamma^* \pi\omega$ vertex function in the space-like region which, at small $t$, 
can be regarded as the $\pi\omega$ transition form factor. The extracted information 
is complementary to that on the form factor in the time-like region derived from data 
\ci{sns,cleo,belle} on electron-positron annihilation into $\pi^0 \omega$, see 
\ci{roig14} for a recent analysis.

A particular combination of $\omega$ SDMEs isolates the so-called unnatural-parity ($U$) 
contribution to the $\gamma^*p\to \omega p$ cross section which, for the kinematics
of the HERMES experiment, is strongly dominated by pion exchange. The $\pi\omega$ 
form factor can easily be extracted from this combination of SDMEs provided the
natural-parity ($N$) contribution to the cross section is known. Since this is not the 
case experimentally at present we are forced to rely on our detailed analysis of exclusive 
meson leptoproduction at small $\xbj$ within the handbag approach \ci{GK3,GK5,GK6}. In 
combination with results of a GPD analysis of the electromagnetic form factors 
\ci{DFJK4,DK13}, we have extracted a set of GPDs which allows us to compute the 
natural-parity contribution to the $\gamma^*p\to \omega p$ cross section and, 
subsequently, the $\pi\omega$ transition form factor from the HERMES data on the $\omega$ 
SDMEs \ci{hermes-omega}. There is also a CLAS measurement of these SDMEs \ci{clas-omega}.   
These data are, however, characterized by large $\xbj$, small $W$ and rather large $t$.
Since the set of GPDs we have determined from meson leptoproduction is optimized for small 
$\xbj$ we cannot reliably compute from it the natural-parity contribution to the 
$\gamma^*p\to \omega p$ cross section for the kinematics of the CLAS experiment and 
therefore we are unable to determine the $\pi\omega$ transition form factor from these data.

In the next section we discuss the role of the pion pole in vector-meson leptoproduction.
In Sect.\ 3 we extract the $\pi\omega$ form factor from the HERMES data. Sect.\ 4
is devoted to a discussion of various partial cross sections which can be obtained from
combinations of SDMEs and to a comparison of the experimental results with the theoretical 
ones obtained from the handbag approach in combination with the pion-exchange contribution.
This further probes the $\pi\omega$ form factor extracted in Sect.\ 3. In Sect.\ 5 we 
present our results for the  SDMEs and, in Sect.\ 6, we comment on spin asymmetries. Our 
summary is given in Sect.\ 7.  

\section{The pion pole}
\label{sec:pole}
\begin{figure}
\begin{center}
\includegraphics[width=0.52\tw]{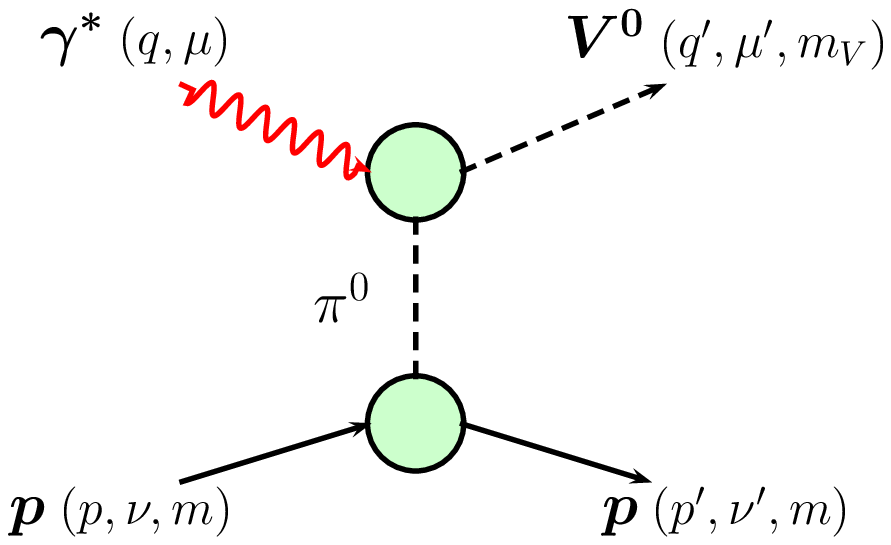} \hspace*{0.03\tw}
\includegraphics[width=0.41\tw]{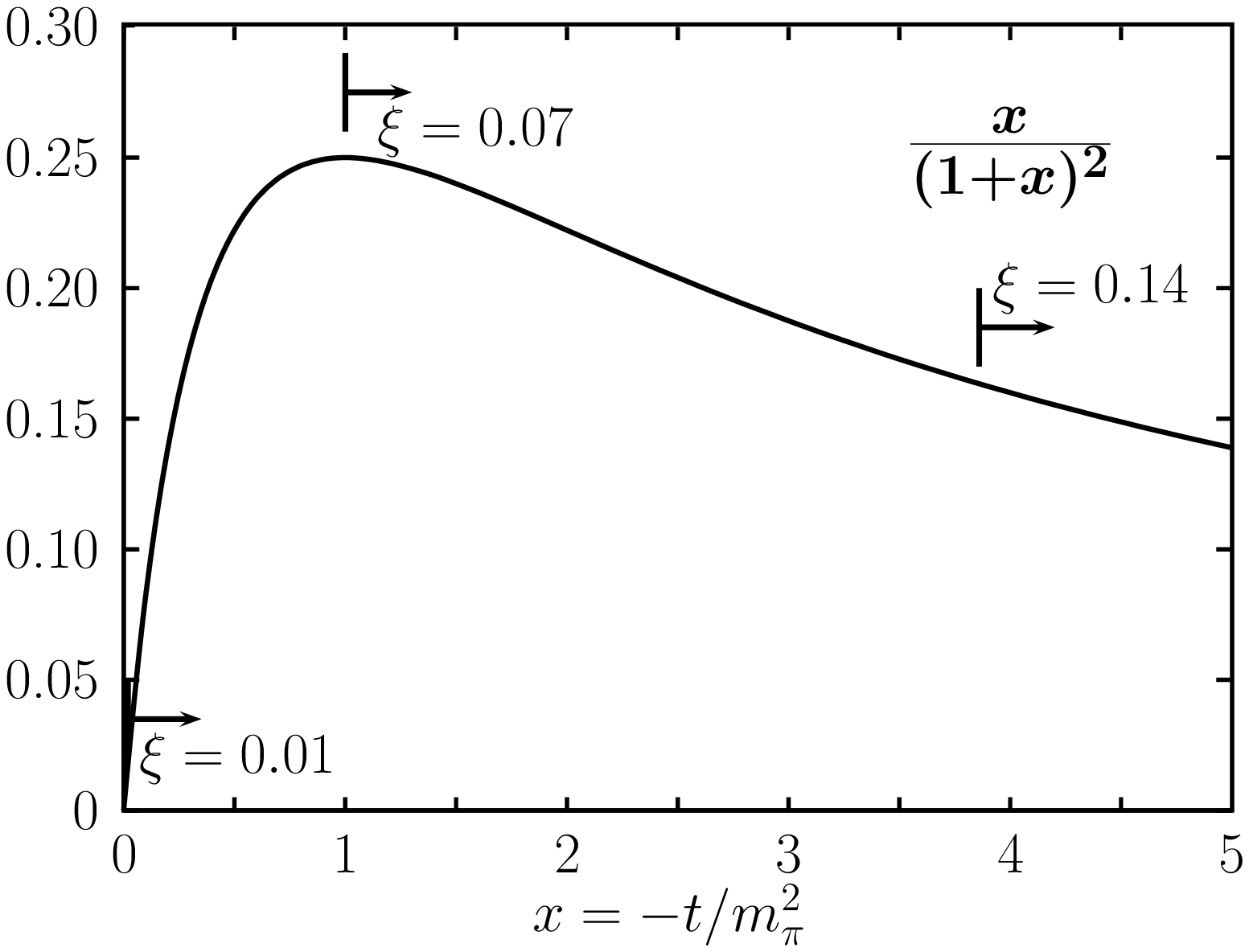}
\end{center}
\caption{Left: The pion-exchange graph in vector-meson leptoproduction. The momenta, 
helicities and masses of the particles are specified in the brackets.
Right: The function $x/(1+x)^2$ where $x=-t/m_\pi^2$. The outset of the physical region 
($t\leq t_0$) is indicated for values of skewness typical for JLab12, HERMES and COMPASS.}
\label{fig:ope}
\end{figure}

Here, in this section, we are going to discuss one-pion exchange in leptoproduction of vector 
mesons ($V=\rho^0, \omega$). The momenta, helicities and masses for the process 
$\gamma^* p\to V p$ are specified in Fig.\ \ref{fig:ope}. In analogy to $\pi^+$ leptoproduction 
\ci{koerner74} one may write the pion-exchange contribution to the helicity amplitudes of 
vector-meson leptoproduction as 
\ba
{\cal M}^{\rm pole}_{\mu'\nu',\mu\nu}&=& i \frac{ g_{\pi NN}F_{\pi NN}(t)}{t-m_\pi^2}
                     \bar{u}(p',\nu')\gamma_5 u(p,\nu) \nn\\
      &\times& \langle V;q',\mu'|j^{\rm el}(0)\cdot \varepsilon_\gamma(\mu)|\pi; q_\pi=q'-q\rangle 
\label{eq:vertex}
\ea
where $g_{\pi NN}$ is the pion-nucleon coupling constant for which we adopt the value 
$13.1\pm 0.2$, and $F_{\pi NN}$ is a form factor that describes the $t$-dependence of the 
pion-nucleon coupling. In concord with \ci{GK6} this form factor is parametrized as
\be
F_{\pi NN} \= \frac{\Lambda_N^2-m_\pi^2}{\Lambda_N^2-t}
\label{eq:piNNFF}
\ee
in the small $-t$ region ($\lsim 0.5\,\gev^2$). For the parameter $\Lambda_N$ we take the value
$(0.44\pm 0.07)\,\gev$ as in \ci{GK6}.  

The current matrix element in \req{eq:vertex} reads
\be 
\langle V;q',\mu'|j^{\rm el}_\kappa(0)|\pi; q_\pi=q'-q\rangle
\= e_0 g_{\gamma^*\pi V}(Q^2,t) \eps_{\kappa\lambda\rho\sigma}q^\lambda\varepsilon^{*\rho}_V(\mu')
                        q'{}^{\sigma}
\label{eq:current}
\ee
where $t=(q'-q)^2$ is the virtuality of the pion. At small $-t$, i.e.\ near the pole, but 
large $Q^2$ one may ignore the $t$-dependence of the vertex function $g_{\gamma^*\pi V^0}$
and regard it as the $\pi V$ transition form factor
\be
   g_{\pi V}(Q^2)\equiv g_{\gamma^*\pi V}(Q^2,m_\pi^2) \simeq g_{\gamma^*\pi V}(Q^2,t) \,.
\label{eq:FFapprox}
\ee 
Chernyak and Zhitnitsky \ci{CZ84} have sketched the calculation of the $\pi V$ transition
form factor within perturbative QCD. This is a complicated task: two- and three-particle
configurations have to be taken into account as well as leading- and higher-twist wave 
functions. In the collinear limit some of the convolutions are infrared singular which, 
according to \ci{CZ84}, may be regularized by Sudakov effects and parton transverse momenta 
in the denominators of the hard propagators. This is precisely the method we have used
in our analysis of meson leptoproduction \ci{GK3,GK5} for the regularization of infrared
singularities occurring in photon-meson transition amplitudes other than 
$\gamma^*_L\to V_L^{\phantom{*}}$ (here and in the following the label $L(T)$ denotes a 
longitudinally (transversely) polarized photon or meson). An outcome of this perturbative 
calculation of the transition form factor is that 
\be
g_{\pi V} \sim 1/Q^4
\ee
at large $Q^2$. Another result is that the ratio of the $\pi\rho^0$ and $\pi\omega$ form factor 
is governed by the quark charges
\be
              g_{\pi\rho} \simeq \frac{e_u+e_d}{e_u-e_d}\,g_{\pi\omega}
\label{eq:charge-ratio}
\ee
leaving aside possible differences in the $\rho$ and $\omega$ wave functions. The
quark charges $e_a$ are given in units of the positron charge $e_0$. We note
that the $1/Q^4$ fall has already been pointed out in \ci{farrar,vainstein}. Chernyak and 
Zhitnitsky estimated the strength of the $\pi\rho$ form factor to amount to
\be
  g_{\pi\rho} \simeq 0.4\,\gev^3/Q^4
\ee
at large $Q^2$. Similar numerical values for this form factor have been obtained from 
light-cone sum rules \ci{braun94,khodjamirian99}. In contrast to \ci{CZ84} only soft
contributions are taken into account in the latter work. 

Experimentally, the transition form factors are unknown in the space-like region except
at $Q^2=0$ where they control the radiative decays of the vector mesons
\be
\Gamma(V\to \pi\gamma) \= \frac1{24} \ale |g_{\pi V}(0)|^2 m_V^3\big[1-m_\pi^2/m_V^2\big]^2
\ee
(see for instance \ci{compilation}). From the branching ratios of the radiative decays of the 
vector mesons and the total decay widths quoted in \ci{pdg} one finds~\footnote{
These values agree with those quoted in \ci{cisek14} if the different normalization in the 
latter work is considered.}
\ba
|g_{\pi\omega}(0)|&=& (2.30\pm 0.04)\,\gev^{-1}\,, \nn\\
|g_{\pi\rho^0}(0)|&=& (0.85\pm 0.06)\,\gev^{-1}\,.
\label{eq:FF(0)}
\ea
Approximately, these values also obey the charge ratio \req{eq:charge-ratio} although the 
radiative decays of the vector mesons are not controlled by perturbative QCD.

The pion-pole contribution to the (light-cone) helicity amplitudes of $\gamma^*p\to Vp$ can 
readily be worked out from \req{eq:vertex} and \req{eq:current}:
\ba
{\cal M}^{\rm pole}_{++,++}&=&-{\cal M}^{\rm pole}_{-+,-+} 
                  \= \frac{\varrho_{\pi V}}{t-m_\pi^2}
                            \,\frac{m\xi Q^2}{\sqrt{1-\xi^2}}\,
                     \Big[1 - \xi^2\, \frac{4m^2-t}{Q^2}\Big]\,,\nn\\
{\cal M}^{\rm pole}_{+-,++}&=&-{\cal M}^{\rm pole}_{--,-+} \ 
            \= -  \frac{\varrho_{\pi V}}{t-m_\pi^2}
                            \,\frac{\sqrt{-t'}Q^2}{2}\, 
                  \Big[1 - \xi^2\, \frac{4m^2-t}{Q^2}\Big]\,,  \nn\\ 
{\cal M}^{\rm pole}_{++,0+}&=& {\cal M}^{\rm pole}_{-+,0+} 
             \= \phantom{-}  \frac{\varrho_{\pi V}}{t-m_\pi^2}
                            \,\sqrt{2}m\xi Q \sqrt{-t'}\,, \nn\\
{\cal M}^{\rm pole}_{+-,0+}&=& {\cal M}^{\rm pole}_{--,0+}  
          \= \phantom{-}\frac{\varrho_{\pi V}}{t-m_\pi^2}
                            \,\sqrt{\frac{1-\xi^2}{2}}t'Q \Big]\,.
\label{eq:hel-ampl}
\ea
Here, the form factors for the coupling of the pion to the vector meson and to the proton 
are combined in the quantity
\be
\varrho_{\pi V}\=e_0 g_{\pi V}(Q^2) g_{\pi NN} F_{\pi NN}(t)\,.
\ee
The skewness, $\xi$, is related to $\xbj$ by
\be
\xi \= \frac{\xbj}{2-\xbj} \Big(1+ \frac{m_V^2}{Q^2}\Big)
\ee
in the photon-proton center-of-mass system specified by $p=\bar{p}-\Delta/2$ and 
$p'=\bar{p}+\Delta/2$ where $\bar{p}=(p+p')/2$ and $\Delta=p'-p$. As usual, $t'=t-t_0$
where 
\be
t_0\=-4m^2\frac{\xi^2}{1-\xi^2}
\ee
is the minimal value of $-t$ attainable in the scattering process $\gamma^*p\to Vp$. The 
contribution of the pion pole to the $\gamma_T^*\to V_T^{\phantom{*}}$
and $\gamma^*_L\to V_T^{\phantom{*}}$ cross sections reads
\ba
\frac{d\sigma^{\rm pole}}{dt}(\gamma^*_T\to V_T^{\phantom{*}})&=&
                         \frac{1}{2\kappa}\,\frac{-t}{(t-m_\pi^2)^2}\,
                Q^4\varrho_{V\pi}^2\big[1-2\xi^2\frac{4m^2-t}{Q^2}\big]\,, \nn\\
 \frac{d\sigma^{\rm pole}}{dt}(\gamma^*_L\to V_T^{\phantom{*}})&=& \frac{1}{\kappa}\,
                  \frac{tt'}{(t-m_\pi^2)^2}\,
                     (1-\xi^2)Q^2\varrho_{V\pi}^2\,,
\label{eq:pole-cross-section}
\ea
where $\kappa$ denotes the phase-space factor and $\Lambda$ the familiar triangle
function
\be
\kappa\=16\pi(W^2-m^2)\sqrt{\Lambda(W^2,-Q^2,m^2)}\,.
\ee
With regard to the large $Q^2$ behavior of the $\pi V$ form factor these cross sections 
are suppressed by $1/Q^2$ and $1/Q^4$ as compared to the asymptotically leading 
$\gamma_L^*\to V_L^{\phantom{*}}$ one which is known to fall as $1/Q^6$ at fixed $\xbj$ (or $\xi$) 
\ci{diehl03}. The pion-pole contribution to the $\gamma^*_T\to V_{-T}^{\phantom{*}}$ and 
$\gamma^*_T\to V_L^{\phantom{*}}$ cross sections are even stronger suppressed and therefore 
neglected in this work. Its contribution to the $\gamma^*_L\to V_L^{\phantom{*}}$ cross section 
is strictly zero.

A structure similar to \req{eq:pole-cross-section} is also found for the pion-pole
contribution to the longitudinal cross section of $\pi^+$ leptoproduction \ci{GK5}
\be
\frac{d\sigma^{\rm pole}}{dt}(\gamma^*_L\to \pi^+)\=\frac{1}{\kappa}\,
            \frac{-t}{(t-m_\pi^2)^2}\,Q^2\varrho_{\pi\pi}^2\,.
\label{eq:piplus}
\ee
where $\varrho_{\pi\pi}=\sqrt{2}e_0F_\pi(Q^2)g_{\pi NN}F_{\pi NN}(t)$ and $F_\pi(Q^2)$
being the electromagnetic form factor of the pion. Characteristic of pion exchange is 
the factor $-t/(t-m_\pi^2)^2$ in \req{eq:pole-cross-section} and \req{eq:piplus} with a 
zero at $t=0$ and a maximum at $t=-m_\pi^2$, see Fig.\ \ref{fig:ope}. This factor 
dominates the $t$-dependence of the relevant cross sections at small $-t$. Depending on 
the value of the skewness the maximum of this factor and, hence, of the cross sections, 
lies in or out of the physical scattering region. 

A reliable extraction of the $\pi V$ transition form factor or the electromagnetic 
form factor of the pion from experiment necessitates data as close as possible to the 
position of the pion pole in order to see the characteristic $t$-dependence
of pion exchange and to justify the neglect of the virtuality of the exchanged pion
at the $\gamma^*\pi V$ vertex. This can only be achieved for sufficiently small skewness.
The extraction of the transition form factor at large $Q^2$ therefore requires 
large $\gamma^*p$ c.m.s.\ energy, $W$, with the consequence of a small pion-pole 
contribution. A reasonable compromise seems to be an energy $W \simeq 3-8\,\gev$ for a 
measurement of the form factor in the range $Q^2=2 - 5\,\gev^2$. This is the case for 
the HERMES experiment \ci{hermes-omega} and can perhaps be realized at the upgraded JLab.

Finally, we want to remark that one may use a reggeized version of pion exchange.
However, owing to the Goldstone-boson nature of the pion the corresponding Regge
trajectory, fixed by the pion and the $\pi(1670)$, is rather flat.
Since we are only interested in very small values of $t$ and a rather 
narrow range of energy both variants of pion exchange differ only slightly.

\section{Extraction of the $\pi\omega$  transition form factor}
For a reliable extraction of the $\pi\omega$ or the pion electromagnetic form factor
it is important to reduce the background, i.e.\ contributions from other dynamical 
mechanisms. In the case of the electromagnetic form factor of the pion this is 
achieved by the familiar longitudinal/transverse separation of the cross section for 
$\pi^+$ production. For the case at hand, the isolation of the unnatural-parity 
cross section ensures the background suppression. The natural/unnatural parity 
separation can be accomplished with the help of a particular combination of SDMEs
\ba
U_1&=& 1- r_{00}^{04} + 2 r_{1-1}^{04} - 2r_{11}^1 -2 r_{1-1}^1 \nn\\
  &=& \frac{2}{\kappa d\sigma/dt} \sum_{\nu'}\left[ |{\cal M}^U_{+\nu',++}|^2 
             +2\varepsilon |{\cal M}^U_{+\nu',0+}|^2\right]\,.
\label{eq:u1}
\ea
The normalization of $U_1$ is related to the transverse and longitudinal differential 
cross sections by $d\sigma/dt=d\sigma_T/dt + \varepsilon d\sigma_L/dt$
where $\varepsilon$ is the ratio of the longitudinal and transverse photon fluxes 
(for HERMES kinematics: $\varepsilon\simeq 0.8$). The helicity amplitudes occurring 
in \req{eq:u1} are their unnatural-parity parts, defined by
\be
{\cal M}^U_{\mu'\nu',\mu\nu}\=\frac12 \Big[{\cal M}_{\mu'\nu',\mu\nu} - (-1)^{\mu-\mu'}
                         {\cal M}_{-\mu'\nu',-\mu\nu}\Big]\,.
\label{eq:unnatural}
\ee
The natural-parity parts of the helicity amplitudes are analogously defined:
\be
{\cal M}^N_{\mu'\nu',\mu\nu}\=\frac12 \Big[{\cal M}_{\mu'\nu',\mu\nu} + (-1)^{\mu-\mu'}
                         {\cal M}_{-\mu'\nu',-\mu\nu}\Big]\,.
\label{eq:natural}
\ee
The $N$- and $U$-type amplitudes possess the property
\ba
{\cal M}_{-\mu'\nu',-\mu\nu}^N &=& \phantom{-}(-1)^{\mu-\mu'}{\cal M}_{\mu'\nu',\mu\nu}^N\,, \nn\\
{\cal M}_{-\mu' \nu',- \mu\nu}^U&=& -(-1)^{\mu-\mu'}{\cal M}_{\mu'\nu',\mu\nu}^U\,.
\ea
The exchange of a particle of either natural or unnatural parity leads to such a 
behavior (see the pion-exchange amplitude \req{eq:hel-ampl}). In the differential 
cross sections there is no interference between the natural- and unnatural-parity
amplitudes.
 
For the SDMEs we use the notation introduced by Schilling and Wolf \ci{schilling}.
In the relations of the SDMEs to the helicity amplitudes we take care
of the conventional normalization of the polarization vector for longitudinal
virtual photons: $\eps_\gamma(0)\cdot \eps_\gamma(0)=1$, c.f.\ the discussion
in \ci{diehl07}. This convention has already been used for the pion-pole
amplitudes \req{eq:hel-ampl}.

Since the unseparated cross section for $\omega$ production has not been measured
by the HERMES collaboration we have to take care of its unnatural-parity part. 
Consequently, we rewrite
\req{eq:u1} as
\ba
\sum_{\nu'}\left[|{\cal M}^U_{+\nu',++}|^2 +2\varepsilon |{\cal M}^U_{+\nu',0+}|^2\right]
   &=& \frac{U_1}{2-U_1}\left\{\sum_{\nu'}\left[|{\cal M}^N_{+\nu',++}|^2 
         +\varepsilon |{\cal M}^N_{0\nu',0+}|^2\right] \right. \nn\\
   &+&\left.  |{\cal M}^N_{0+,++}|^2 +\frac12|{\cal M}_{0-,++}|^2\right\}\,.
\label{eq:separation}
\ea
The left-hand side of this equation, dominated by the pion-pole contribution 
\req{eq:hel-ampl}, includes terms quadratic and, provided there is a background to these 
amplitudes, linear in the form factor $g_{\pi\omega}$. For this possible background 
as well as for the amplitudes on the right-hand side of \req{eq:separation} we exploit
results from our previous studies of hard exclusive meson production at small skewness
within the handbag approach \ci{GK3,GK5,GK6} in which the helicity amplitudes are given by 
convolutions of hard subprocess amplitudes and GPDs. The partonic subprocess has been 
calculated by us within the modified perturbative approach in which quark transverse
momenta and Sudakov suppressions are taken into account. From that analysis we have
extracted a set of GPDs for gluons and quarks which include $H$, $E$ and $\widetilde H$ 
as well as the transversity GPDs $\bar{E}_T=2\widetilde{H}_T+E_T$ and $H_T$. The GPD 
$\widetilde E$ has been neglected. In principle, pion exchange contributes to $\widetilde E$ 
\ci{man98,goeke99}. We however replace this contribution by the one-particle exchange term 
discussed in the preceding section (see also next section). The parametrizations of the 
GPDs are specified in \ci{GK3,GK6}. The convolutions of $H$, $E$ (contributing 
to $\gamma^*_L\to V_L^{\phantom{*}}$ and $\gamma^*_T\to V_T^{\phantom{*}}$ transition amplitudes)  
and $\bar{E}_T$ (contributing to $\gamma^*_T\to V_L^{\phantom{*}}$ transitions) behave like 
natural-parity exchanges \req{eq:natural}, those of $\widetilde H$ (feeding the 
$\gamma^*_T\to V_T^{\phantom{*}}$ amplitudes) like unnatural parity \req{eq:unnatural}. A 
special case is the convolution of $H_T$ which fix the helicity non-flip amplitude 
${\cal M}_{0-,++}$. Its parity counterpart, ${\cal M}_{0-,-+}$, is suppressed by $t/Q^2$. 
Hence, the convolution of $H_T$ does not have a definite parity \ci{GK6,diehl01}. No parity 
label is therefore assigned to the amplitude ${\cal M}_{0-,++}$ in \req{eq:separation}.
The $\gamma_L^*\to V_T^{\phantom{*}}$ and $\gamma_T^*\to V_{-T}^{\phantom{*}}$ transition 
amplitudes are assumed to be zero except of the pion-exchange contribution to the first 
ones~\footnote{A non-zero $\gamma^*_T\to V_{-T}^{\phantom{*}}$ amplitude could be generated 
by gluon transversity \ci{GK7}.}. We stress that our handbag approach is designed for 
$\xi\lsim 0.1$,  $-t'<0.7\,\gev^2$, $W\gsim 4\,\gev$, and $Q^2\gsim 2\,\gev^2$. 

\begin{figure}[t]
\begin{center}
\includegraphics[width=0.45\tw]{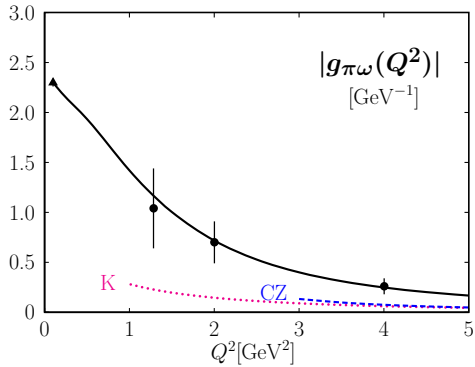}
\end{center}
\caption{The $\pi\omega$ transition form factor versus $Q^2$ extracted 
from the HERMES data \ci{hermes-omega} on $U_1$ (at $W=4.8\,\gev$ and $t'=-0.08\,\gev^2$). 
The value of the form factor at $Q^2=0$, represented by a triangle, is shifted for the 
ease of legibility. The solid line represents the parametrization \req{eq:FFparametrization}; 
the dashed (dotted) one the pQCD (light-cone sum rule) result quoted in \ci{CZ84}
(\ci{khodjamirian99}) for the $\pi\rho$ form factor multiplied by 3.}
\label{fig:FF}
\end{figure}

The SDME combination $U_1$ appearing in \req{eq:separation} is known from the HERMES measurement 
\ci{hermes-omega}. The natural-parity amplitudes in \req{eq:separation} as well as a background 
to the amplitude ${\cal M}^U_{++,++}$ is evaluated from the above described set of GPDs along 
the lines we computed $\rho^0$ production \ci{GK3}. For this computation we apply the same 
wave functions for the $\omega$ as for the $\rho^0$ except that the decay constants of the 
$\rho^0$ are replaced by those of the $\omega$ ($f_\omega=187\,\mev$ and $149\,\mev$ for 
longitudinally and transversely polarized $\omega$-mesons, respectively). It turns out that the 
contribution of $\widetilde H$ to ${\cal M}^U_{++,++}$ is dominantly imaginary at small $-t$ 
with the consequence of a very small interference with the pion-exchange amplitudes. Therefore, 
the two solutions of \req{eq:separation} for $g_{\pi\omega}$ differ only by the sign within errors.
In other words, we are only able to extract the absolute value of the form factor from $U_1$ 
for which we take the average of the two solutions. Our results for the form factor are shown 
in Fig.\ \ref{fig:FF}. We also quote a result for the form factor at $Q^2=1.284\,\gev^2$ for 
which HERMES has measured the $\omega$ SDMEs too. This form factor value is to be taken with a 
grain of salt since the handbag amplitudes are rather uncertain at such low values of $Q^2$ 
since they are not probed against experiment. Our results for the form factor are apparently 
consistent with the $Q^2=0$ value. The $Q^2$ dependence of the results can be parametrized as
\be
|g_{\pi\omega}(Q^2)| \= \frac{2.3\,\gev^{-1}}{1+Q^2/a^2_1+Q^4/a^4_2}
\label{eq:FFparametrization}
\ee
with $a_1=2.7\,\gev$ and $a_2=1.2\,\gev$. This parametrization will be used if the form factor
is needed at values of $Q^2$ other than 2 or $4\,\gev^2$. For comparison we have also shown in 
Fig.\ \ref{fig:FF} the perturbative and the light-cone sum rule results for the $\pi\rho$ form 
factor \ci{CZ84} multiplied by the charge ratio \req{eq:charge-ratio}. They are substantially 
smaller than our results extracted from the HERMES data on the $\omega$ SDMEs. Our error 
assessment does not only include the experimental errors of $U_1$ but also the uncertainties of 
the handbag amplitudes as well as those of the pion-nucleon coupling. The errors would be 
considerably smaller if the $\omega$ cross section were known from experiment with an error 
smaller than, say, $10\%$. In the absence of such data our results for the $\pi\omega$ form 
factor evidently depend on the model used for the natural-parity amplitudes and one may wonder 
whether we have merely extracted an effective parameter. Nevertheless, the consistency of the 
entire approach supports its interpretation  as an realistic estimate of the $\pi\omega$ 
transition form factor.

The combination $U_1$ for $\omega$ production has also been measured by the CLAS collaboration
\ci{clas-omega}. Its values are about as large as the HERMES data. However, since for the CLAS 
data $W$ lies in the range $1.8 - 2.5\,\gev$ while the range of $Q^2$ is similar to that of the 
HERMES data, $-t$ is larger than $0.5\,\gev^2$ in this experiment. In this kinematical 
situation neither the parametrization of the pion exchange (see \req{eq:piNNFF}, 
\req{eq:FFapprox}) nor the natural-parity handbag amplitudes are reliably known which prevents 
an estimate of the $\pi\omega$ form factor. We will return to this issue in the next section.  
\section{Partial cross sections}
\label{sec:partial}
With the $\pi\omega$ form factor at disposal we are in the position to
evaluate various partial cross section and SDMEs and to compare the results with
the HERMES data \ci{hermes-omega}. The partial cross sections we are using, are defined 
by
\be
\frac{d\sigma(\gamma^*_i\to V_j)}{dt}=\frac1{\kappa s_i}\sum_{\mu_j\nu'\mu_i}
        \big|{\cal M}_{\mu_j\nu',\mu_i+}\big|^2
\nn
\ee
where $i$ and $j$ being either $L$ or $T$ and  $\mu_{L(T)}=0(\pm 1)$. The statistical factor 
is $s_L=1$, $s_T=2$. Neglecting $\gamma^*_T\to V_{-T}^{\phantom{*}}$ transitions and the 
amplitude ${\cal M}_{0-,-+}$, we can decompose the full differential cross section for the 
process $\gamma^* p\to Vp$ into
\ba
\frac{d\sigma}{dt}&=&\frac{d\sigma^N}{dt}(\gamma_T^*\to V_T^{\phantom{*}}) + 
   \frac{d\sigma^U}{dt}(\gamma_T^*\to V_T^{\phantom{*}})
                      +\frac{d\sigma}{dt}(\gamma_T^*\to V_L^{\phantom{*}})\nn\\
  && + \varepsilon \frac{d\sigma^N}{dt}(\gamma_L^*\to V_L^{\phantom{*}})
   + \varepsilon \frac{d\sigma^U}{dt}(\gamma_L^*\to V_T^{\phantom{*}})\,.
\ea

\begin{figure}
\begin{center}
\includegraphics[width=0.45\tw]{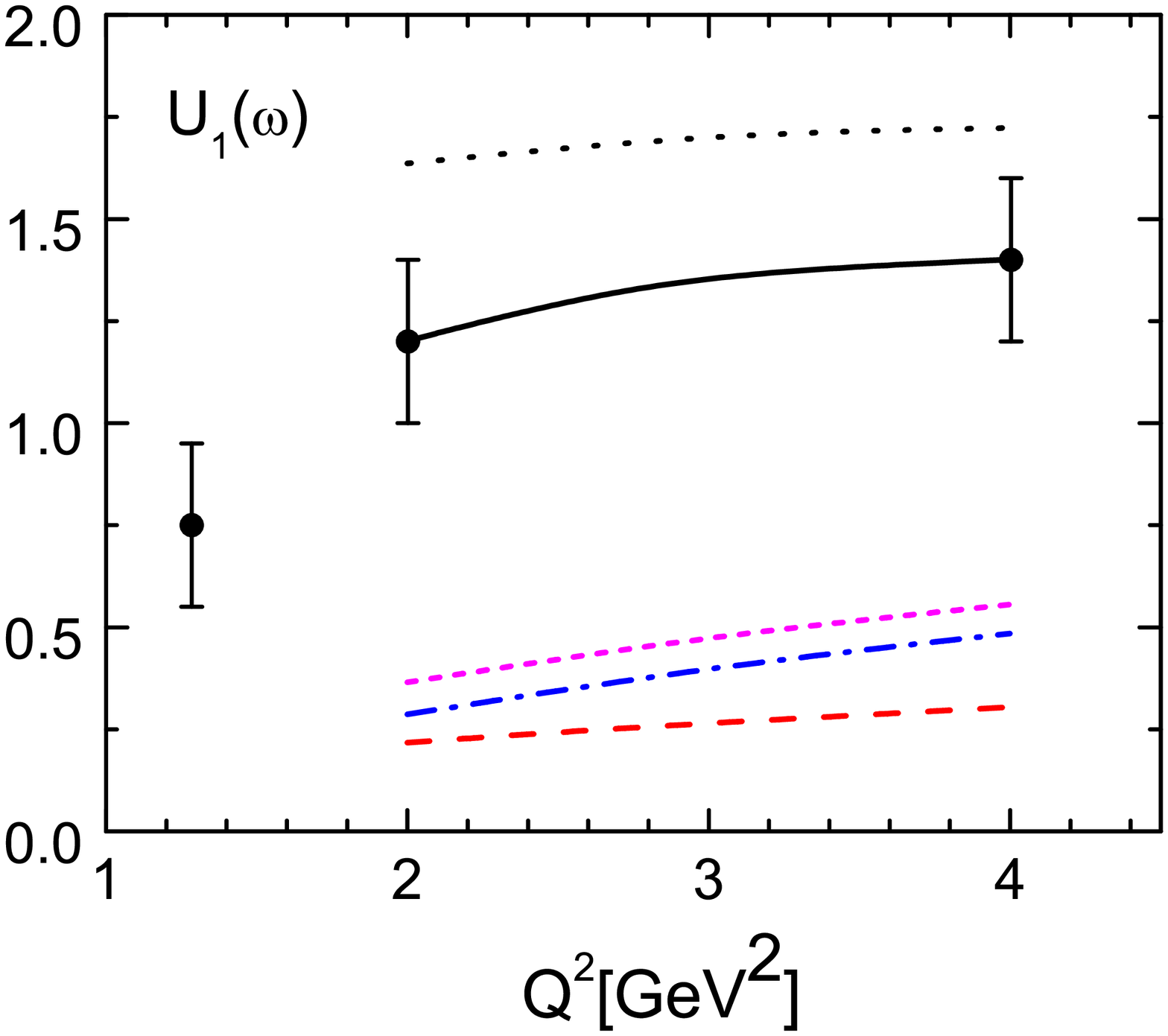}\hspace*{0.05\tw}
\includegraphics[width=0.46\tw]{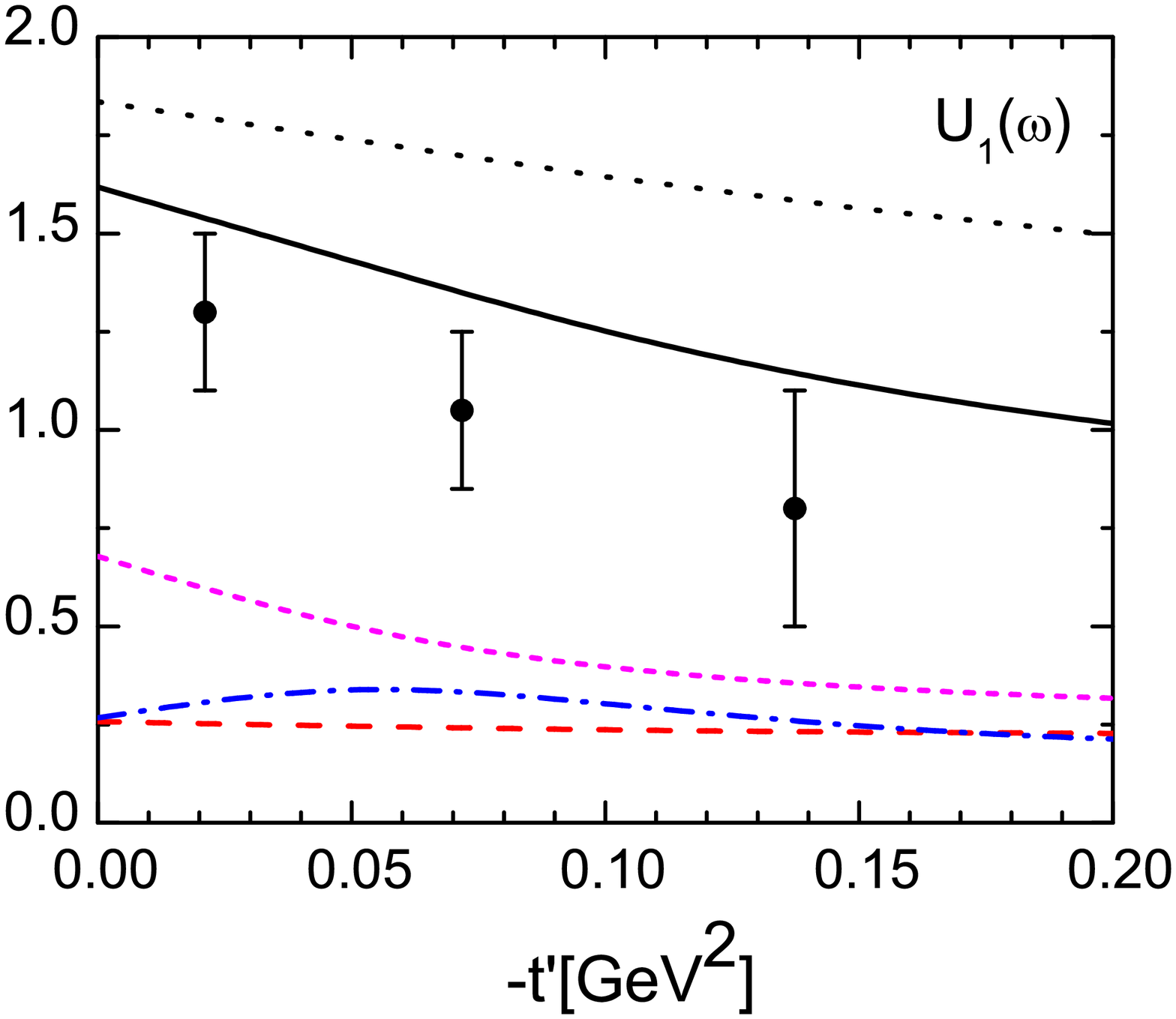}
\end{center}
\caption{Left: $U_1$ versus $Q^2$ for $\omega$ production at $W=4.8\,\gev$ and 
$t'=-0.08\,\gev^2$. Right: $U_1$ versus $t'$ at $W=4.8\,\gev$ and $Q^2=2.42\,\gev^2$. The 
data are taken from \ci{hermes-omega}. The solid (long-dashed) lines represent our results 
from the handbag approach with (without) the pion pole. The dotted and dashed-dotted lines 
are predictions at $W=3.5$ ($\varepsilon=0.7$) and $8\,\gev$ ($\varepsilon=0.96$), 
respectively. The short-dashed line is evaluated from ${\widetilde E}_{\rm pole}$, see text.}
\label{fig:U1}
\begin{center}
\includegraphics[width=0.45\tw]{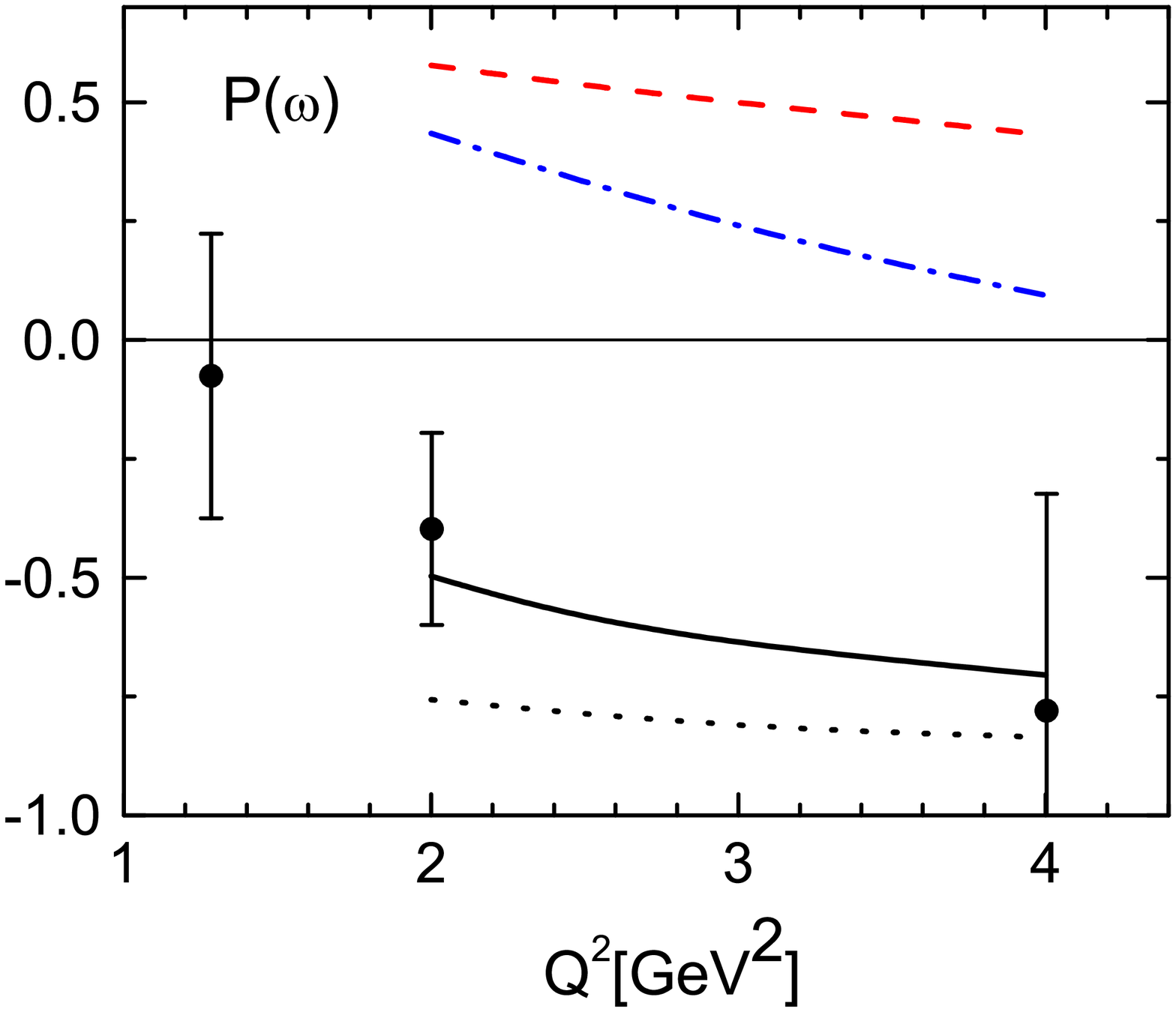}\hspace*{0.03\tw}
\includegraphics[width=0.47\tw]{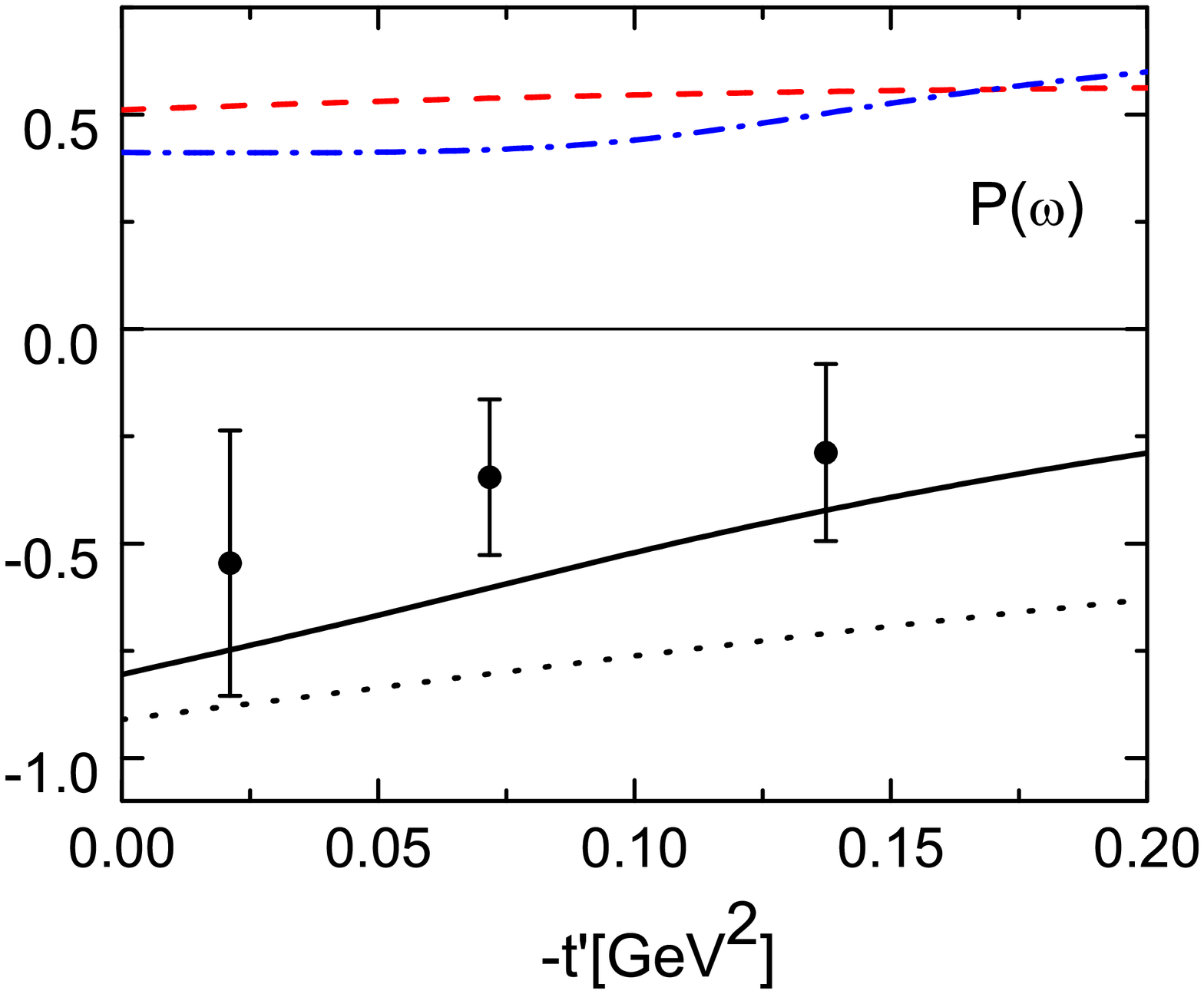}
\end{center}
\caption{$P$ versus $Q^2$ (left) and $t'$ (right). For other notations it is 
referred to Fig.\ \ref{fig:U1}.}
\label{fig:P}
\end{figure}

To begin with we show the $Q^2$- and $t$-dependence of the quantity $U_1$ in Fig.\ \ref{fig:U1}. 
It measures the cross section ratio (c.f.\ \req{eq:u1})
\be
U_1\=2\,\frac{d\sigma^U(\gamma^*_T\to V_T^{\phantom{*}})
             +\varepsilon d\sigma^U(\gamma^*_L\to V_T^{\phantom{*}})}{d\sigma}\,.
\label{eq:u1-diff}
\ee
The $Q^2$-dependence, now evaluated from the interpolation \req{eq:FFparametrization}, 
is of course perfectly reproduced since we extracted the $\pi\omega$ form factor from 
this quantity. The $t$-dependence of our results for $\omega$ production is a bit too 
mild as compared to the HERMES data. We stress that the $t$-dependent HERMES data are
an average of all data for $Q^2> 1\,\gev^2$ \ci{hermes-omega}. Since the trends of our
results towards low $Q^2$ ($<2\,\gev^2$) are in general in reasonable agreement with
experiment we do not think that the inclusion of the low $Q^2$ data in the averages 
is the source of the rather flat $t$-dependence of our results. The reasonable value 
of the $\pi\omega$ form factor at  $Q^2> 1.284\,\gev^2$ (see Fig.\ \ref{fig:FF}) 
supports this supposition.

In order to demonstrate the importance of the pion 
pole we also show results in Fig.\ \ref{fig:U1} for which the pion pole is disregarded. 
In this case $U_1$ is dramatically reduced since the unnatural-parity amplitudes are 
only fed by the GPD $\widetilde H$. The $t$ dependence is much flatter without the 
pion-pole contribution, i.e.\ the differential cross sections appearing in \req{eq:u1-diff} 
have similar $t$ dependencies except of the pion-pole contribution. Predictions for $U_1$ 
at $W=3.5$ and $8\,\gev$ are also displayed in Fig.\ \ref{fig:U1}. A strong energy 
dependence of the pion pole contribution is to be noticed. It is very large at 
$W=3.5\,\gev$ but small at $8\,\gev$. In the latter case the results are close to those
without the pion pole. Note the maximum of $U_1$ at $t'\simeq -0.05\,\gev^2$ and
$W=8\,\gev$ which is a consequence of the factor $-t/(t-m_\pi^2)^2$ discussed in 
Sect.\ \ref{sec:pole}, see Fig.\  \ref{fig:ope}.

It is well-known that the pion pole contributes to the GPD $\widetilde E$ 
\ci{man98,goeke99}:
\be
\widetilde{E}^u_{\rm pole}\=-\widetilde{E}^d_{\rm pole}\=-\Theta(|x|\leq \xi) 
            \frac{mf_\pi}{2\xi}\frac{g_{\pi NN}F_{\pi NN}(t)}{t-m_\pi^2}
                           \Phi_\pi(\frac{x+\xi}{2\xi})
\label{eq:Etilde-pole}
\ee
where $f_\pi$ is the decay constant and $\Phi_\pi$ the distribution amplitude of 
the pion. Evidently, the charge factor between $\omega$ and $\rho^0$ production 
quoted in \req{eq:charge-ratio} immediately follows from \req{eq:Etilde-pole} since 
the combinations $e_u\widetilde{E}^u + e_d\widetilde{E}^d$ and 
$e_u\widetilde{E}^u - e_d\widetilde{E}^d$ contribute to $\omega$ and $\rho^0$ 
production, respectively. It is well-known that an evaluation of the pion-pole 
contribution to $\pi^+$ leptoproduction from ${\widetilde E}_{\rm pole}$ underestimates 
it by far since this way only the one-gluon exchange contribution to the  
electromagnetic form factor of the pion  in \req{eq:piplus} is taken into account 
which is known to amount to only $30-50\%$ of its experimental value \ci{blok08}. 
To probe the contribution from ${\widetilde E}_{\rm pole}$ in $\omega$ leptoproduction 
we evaluate it from the handbag graphs along the same lines as we did it for 
$\widetilde H$, see \ci{GK3}. Since ${\widetilde E}_{\rm pole}$ contributes to the 
$\gamma^*_T\to \omega^{\phantom{*}}_T$ transition amplitude while quark and antiquark 
forming the meson, possess opposite helicities one unit of orbital angular momentum 
is required in the $\omega$ wave function which is represented by a factor 
${\bf k_\perp}$ \ci{GK3,BKK}. A result similar to that for  $\pi^+$ production is 
obtained -  the pion pole contribution to $\omega$ production evaluated through 
\req{eq:Etilde-pole}, is about a factor of 3 smaller than the one-particle 
exchange contribution, see Sect.\ \ref{sec:pole}. This also holds true for $\rho^0$ 
production which motivated us to neglect the pion-pole contribution in our previous 
analysis of this process \ci{GK3}.

Another way to separate the natural- and unnatural-parity cross section is the 
combination~\footnote{In \ci{hermes-omega} a slightly different definition
of $P$ is used.} 
\be
P \= \frac{2r_{1-1}^1}{1-r_{00}^{04} -2 r^{04}_{1-1}} \=
    \frac{d\sigma^N(\gamma^*_T\to V_T^{\phantom{*}}) - d\sigma^U(\gamma^*_T\to V_T^{\phantom{*}})}
          {d\sigma^N(\gamma^*_T\to V_T^{\phantom{*}}) + d\sigma^U(\gamma^*_T\to V_T^{\phantom{*}})}\,.
\ee
Our results for $P$ in $\omega$ production are shown in Fig.\ \ref{fig:P} and compared to 
the HERMES data. Good agreement with experiment is to be noticed as well as the prominent 
role of the pion pole. With the pole the unnatural-parity $\gamma^*_T\to V_T^{\phantom{*}}$ 
cross section is about three times larger than the corresponding natural one at $W=4.8\,\gev$, 
without it the ratio of the two cross sections only amounts to about 1/3. In the latter case 
the unnatural-parity cross section is solely fed by the $\widetilde H$ contribution. The $N/U$ 
ratio strongly depends on the energy. 

\begin{figure}
\begin{center}
\includegraphics[width=0.45\tw]{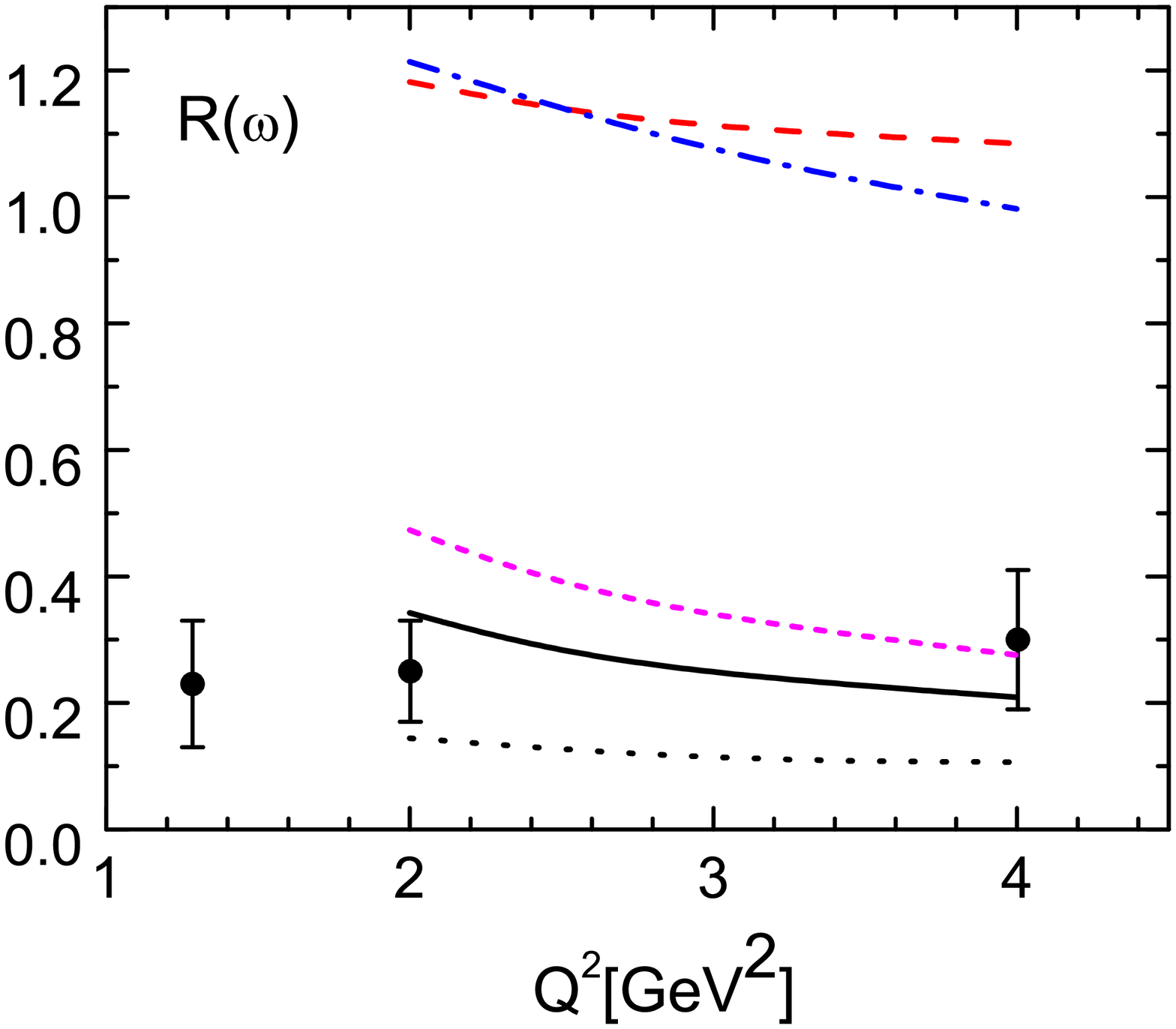}\hspace*{0.05\tw}
\includegraphics[width=0.47\tw]{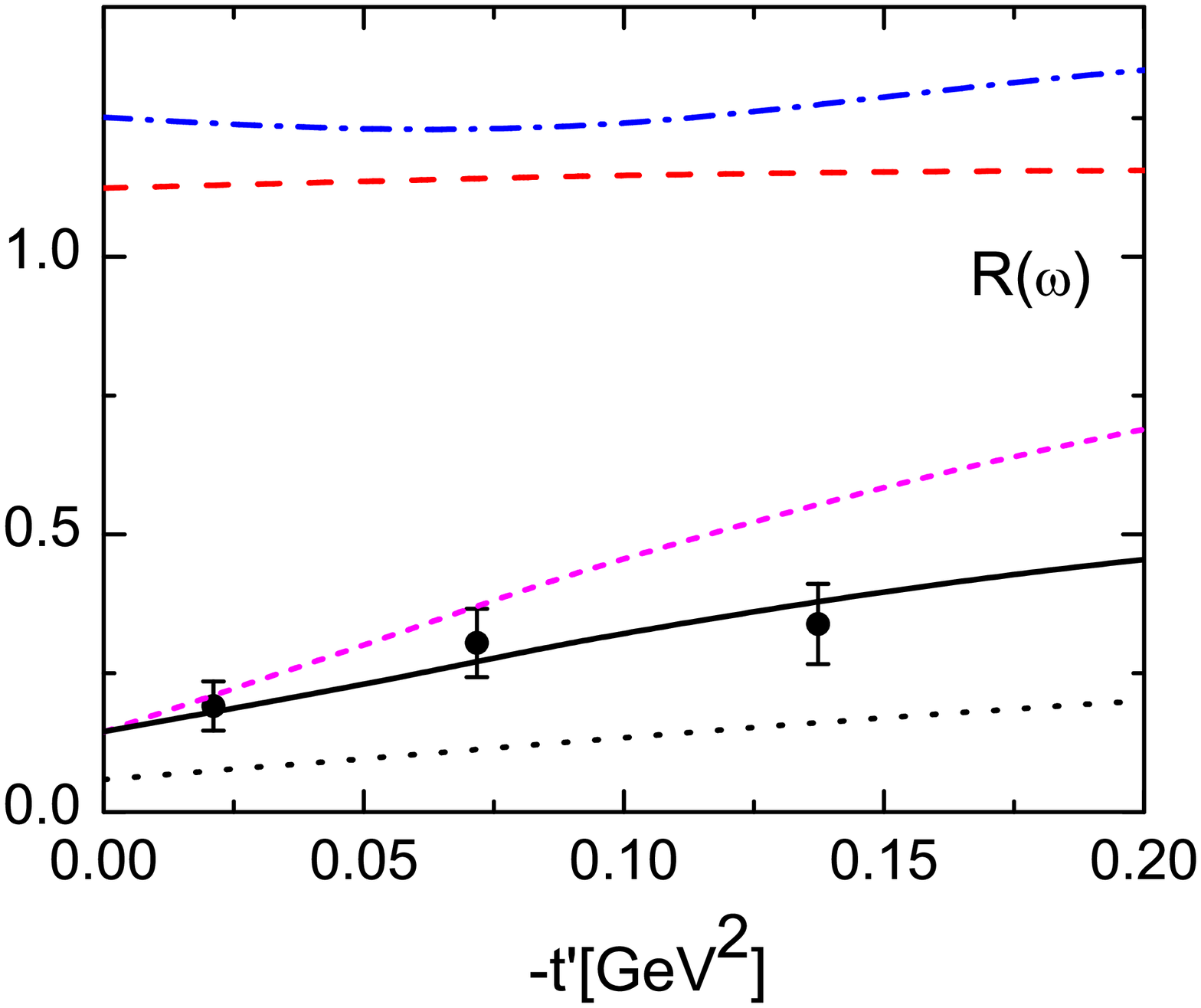}
\end{center}
\caption{$R$ versus $Q^2$ (left) and versus $t'$ (right). For other notations 
it is referred to Fig.\ \ref{fig:U1}. The short-dashed line represents the cross section
ratio for longitudinal and transverse photons.}
\label{fig:R}
\begin{center}
\includegraphics[width=0.45\tw]{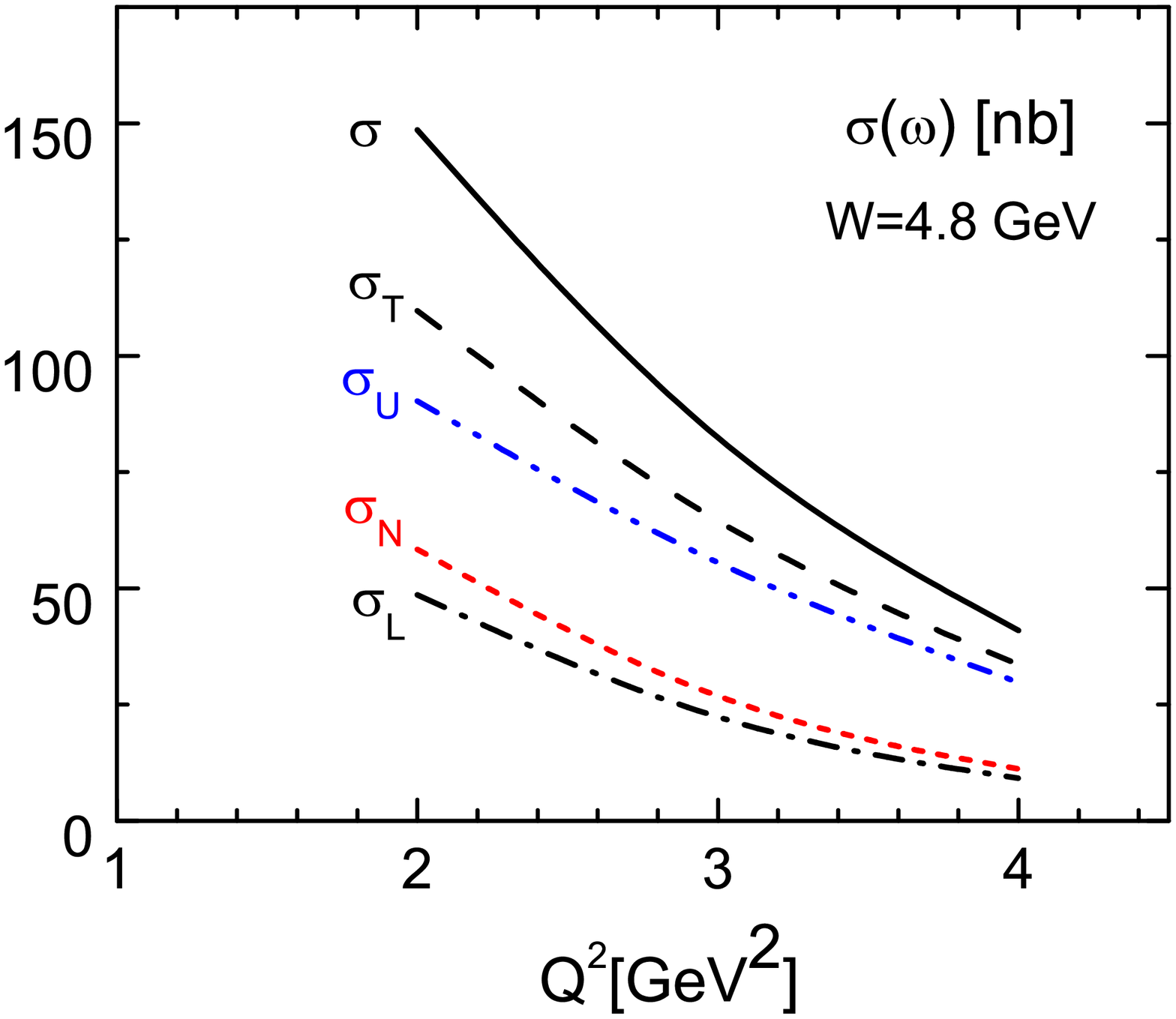}\hspace*{0.05\tw}
\includegraphics[width=0.44\tw]{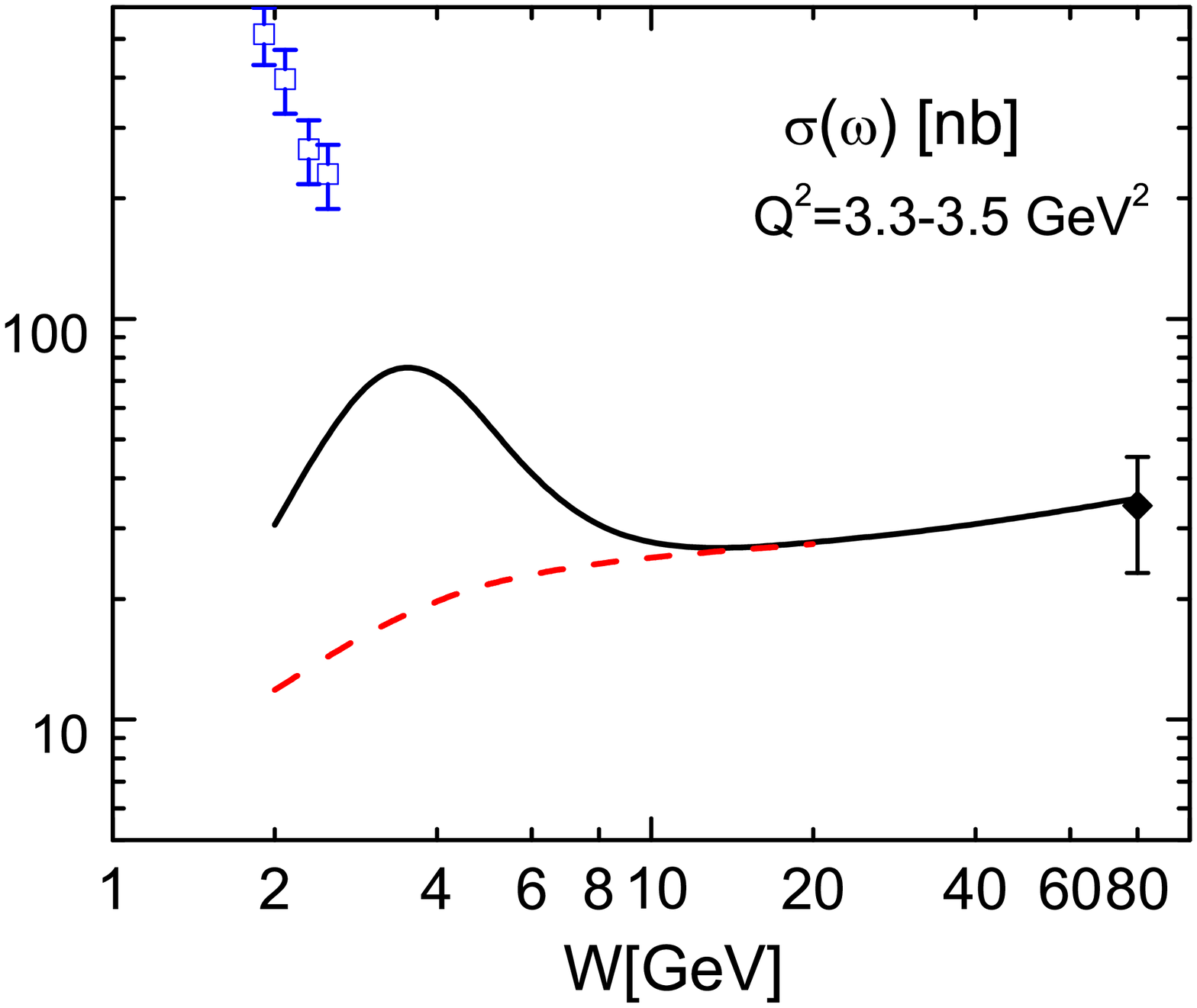}
\end{center}
\caption{Left: The integrated cross sections ($-t'\leq 0.5\,\gev^2$) for longitudinal ($\sigma_L$)
and transverse ($\sigma_T$) photons as well as the natural-parity ($\sigma_N$), 
unnatural-parity ($\sigma_U$) and full ($\sigma$) ones for $\omega$ production versus $Q^2$ 
at $W=4.8\,\gev$. Right: The integrated cross section for $\omega$ production versus $W$ at 
$Q^2=3.3-3.5$. The solid (dashed) line represents our results with (without) the pion pole.
Data are taken from \ci{clas-omega,breitweg}.}
\label{fig:sigma}
\end{figure}

A frequently considered combination is 
\be
R \= \frac1{\varepsilon} \frac{r_{00}^{04}}{1-r_{00}^{04}} 
        \=  \frac{d\sigma(\gamma^*_L\to V_L^{\phantom{*}})
           +\frac1{\varepsilon}d\sigma(\gamma^*_T\to V_L^{\phantom{*}})}
{d\sigma(\gamma^*_T\to V_T^{\phantom{*}}) +\varepsilon d\sigma(\gamma^*_L\to V_T^{\phantom{*}})}\,,
\ee
which is the ratio of the cross sections for longitudinally and transversally polarized
vector mesons. It is often identified with the cross section ratio  for longitudinally and 
transversally polarized photons - $d\sigma_L/d\sigma_T$. This is however only be true if  
the transitions $\gamma^*_L\to V_T^{\phantom{*}}$  and $\gamma^*_T\to V_L^{\phantom{*}}$ are 
strongly suppressed. As can be seen from Fig.\ \ref{fig:R} this is not the case for 
$\omega$ production: $d\sigma_L/d\sigma_T$ is about $30\%$ larger than $R$ at 
$t'=-0.08\,\gev^2$ because of the rather large cross section 
$d\sigma^U(\gamma^*_L\to\omega_T^{\phantom{*}})$. The cross section 
$d\sigma(\gamma^*_T\to \omega_L^{\phantom{*}})$, fed by the transversity GPDs, plays only a 
minor role. Due to the pion-pole contribution which strongly enhances the 
$\gamma^*_T\to \omega_T^{\phantom{*}}$ cross section $R$ is much smaller than 1. This is in sharp 
contrast to $\rho^0$ production where $R$ is somewhat larger than 1 at $Q^2\simeq 3\,\gev^2$ 
and almost independent on energy \ci{hermes-rho,h1}. 

In Fig.\ \ref{fig:sigma} we display various integrated cross sections for $\omega$ production.
As expected  from the discussions of $U_1$, $P$ and $R$ we find $\sigma_T>\sigma_L$ and 
$\sigma_U>\sigma_N$ (the small amplitude ${\cal M}_{0-++}$ is assigned to $\sigma_N$). Neglect
of the pion pole reverses the inequalities. All these features lead us to the conclusion
that, at least for HERMES kinematics, $\omega$ production is very different from the asymptotic 
result of proceeding dominantly through longitudinally polarized photons \ci{collins97}. Our
predictions for the $\omega$ cross section versus $W$ at fixed $Q^2$ ($=3.3-3.5\,\gev^2$) 
shown in Fig.\ \ref{fig:sigma} reveal an interesting behavior: there is a maximum at 
$W\simeq 4\,\gev$ which again - and not surprisingly now - is caused by the pion pole. At 
large $W$, say larger than about $8\,\gev$ its contribution is very small and the $\omega$ 
cross section behaves diffractively like the $\rho^0$ or $\phi$ cross sections \ci{GK3,GK2}, 
i.e.\ slowly increasing with energy. At low values of $W(\lsim 3\,\gev)$ the pion-pole 
contribution \req{eq:pole-cross-section} is decreasing too since with decreasing $W$ but 
fixed $Q^2$ the skewness and, hence, $-t_0$ increase as well. For instance, at $W=2\,\gev$ 
and $Q^2=3.5\,\gev^2$, $t_0$ amounts to $-0.76\,\gev^2$. The cross sections 
\req{eq:pole-cross-section} are therefore only probed in the large $-t$ region in that 
kinematical situation where they fall $\sim 1/t^3$. From Fig.\ \ref{fig:sigma} it is clear that,
for $W\lsim 3\,\gev$ we underestimate the experimental $\omega$ cross section \ci{clas-omega}.
 
\begin{figure}
\begin{center}
\includegraphics[width=0.45\tw]{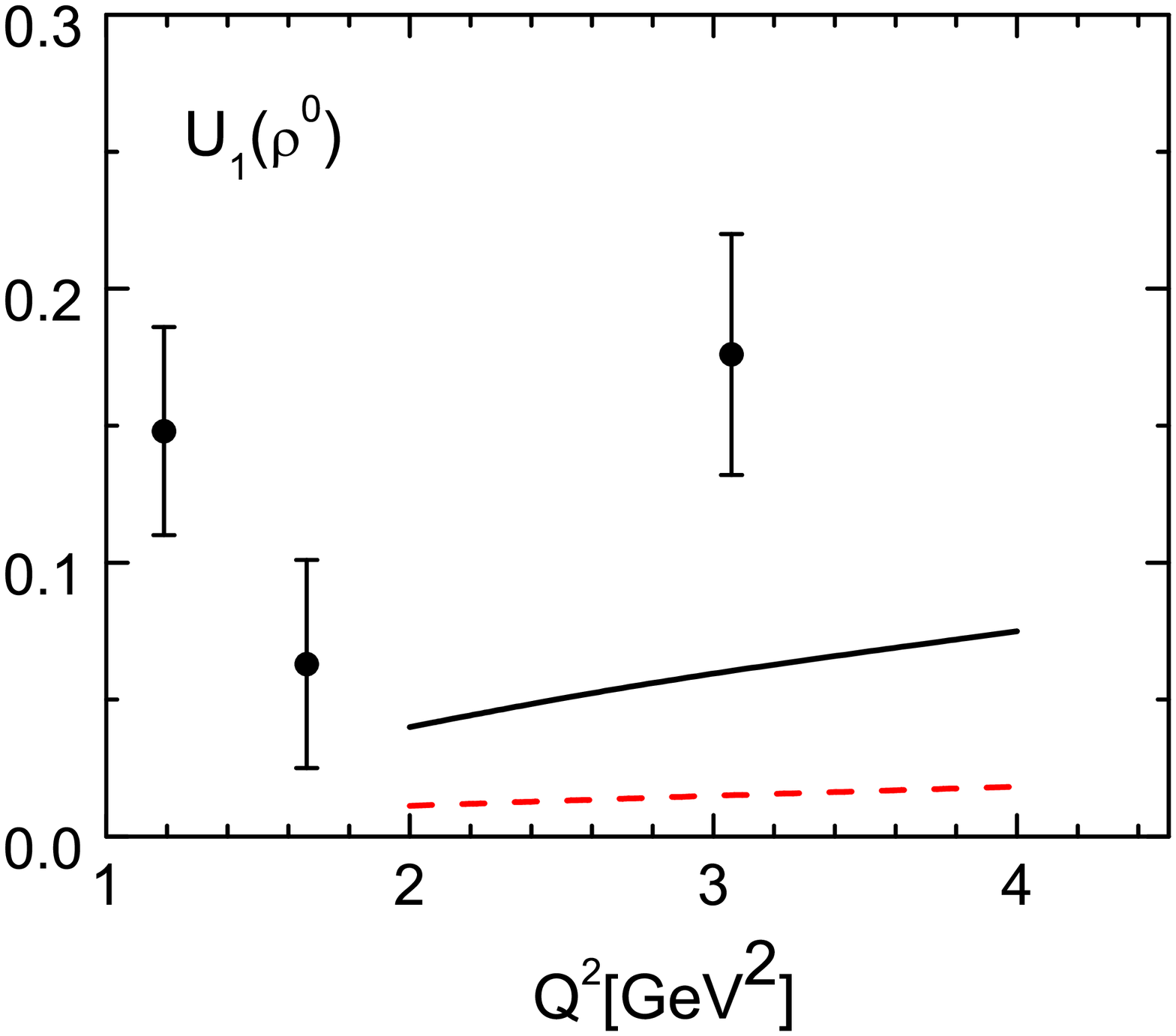}\hspace*{0.05\tw}
\includegraphics[width=0.45\tw]{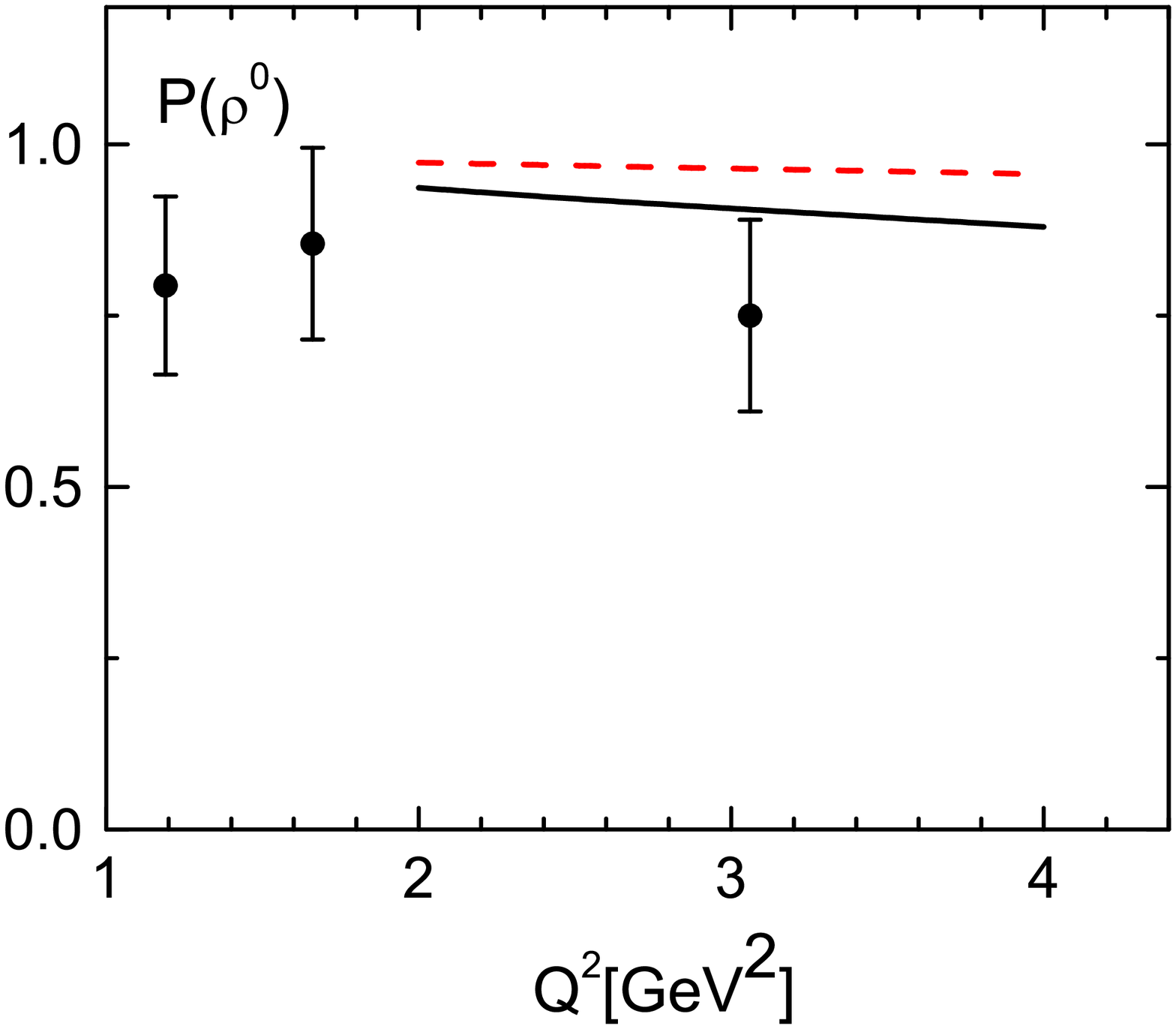}
\end{center}
\caption{$U_1$ (left) and $P$ (right) versus $Q^2$ for $\rho^0$ production at $W=4.8\,\gev$ and 
$t'=-0.13\,\gev^2$. The data are taken from \ci{hermes-rho}. The solid (dashed) lines represent 
our results from the handbag approach with (without) the pion pole.}
\label{fig:U1rho}
\end{figure}

As we mentioned above we have neglected the pion pole in our previous studies of $\rho^0$
leptoproduction \ci{GK3}. Now, treating it as a one-particle exchange instead of evaluating
it from the GPD $\widetilde{E}$ as given in \req{eq:Etilde-pole}, we obtain a larger although
still fairly small effect: According to \req{eq:charge-ratio} and \req{eq:FF(0)}, we have  
$d\sigma^{\rm pole}(\rho^0)\simeq d\sigma^{\rm pole}(\omega)/9$ while for natural-parity,  
one has $d\sigma^N(\rho^0)\simeq 9d\sigma^N(\omega)$ at $W\simeq 5\,\gev$ and 
$Q^2\simeq 3.5\,\gev^2$. Thus, in sharp contrast to $\omega$ production, the pion pole plays 
only a minor role in $\rho^0$ production, it is almost negligible. It increases
the integrated cross section only by about $2\%$ for HERMES kinematics and is similarly
unimportant in most of the other observables. Exceptions are $U_1$ and $P$. As one may see
from Fig.\ \ref{fig:U1rho} the contribution from $\widetilde{H}$ to $U_1$ is tiny, much smaller
than experiment \ci{hermes-rho}. The pion-pole contribution, evaluated from $g_{\pi\omega}/3$
with $g_\pi\omega$ given in \req{eq:FFparametrization}, enhances $U_1$ but the result is 
still somewhat small. Since $P$ is close to 1 (see Fig.\ \ref{fig:U1rho}) the unnatural-parity 
contribution to the $\gamma^*_T\to \rho_T$ transitions is much smaller than the natural-parity 
one. 

Both the cross sections, the $\omega$ and the $\rho^0$ one, decrease by order of magnitude  
between $W=2$ and $4\,\gev$ and are about equal in that region \ci{clas-omega,clas-rho,cornell}. 
This is to be regarded as a hint that the pion pole is not responsible for this decrease. 
Through what it is caused is as yet unknown. Teryaev et al.\ \ci{teryaev} suggest that the 
so-called $D$ term \ci{polyakov-weiss} which we neglect in our parametrization of the GPDs, 
may generate this effect. However, since the sharp drop of the cross section between 2 and 4 GeV 
is also seen in $\rho^+$ production \ci{clas-rhop} but not in the $\phi$ cross section \ci{clas-phi}
this explanation must be taken with care. In any case below about $3\,\gev$ there seems to be 
an additional strong dynamical mechanism beyond our handbag approach and the pion pole.

Through $\omega-\phi$ mixing the pion pole also appears in electroproduction of $\phi$ 
mesons. The $\pi\phi$ transition from factor is related to the $\pi\omega$ one by
\be
g_{\pi\phi}(Q^2)\simeq \sin{(\Phi_V)} g_{\pi\omega}(Q^2)\,.
\ee
The vector-meson mixing angle, $\Phi_V$, in the quark-flavor basis is very small, 
about $3$ degrees, as obtained from the ratio of the $\phi\to \pi\gamma$ and
$\omega\to \pi\gamma$ decay widths \ci{feldmann}. Hence, the neglect of the pion pole 
in $\phi$ production is beyond doubt.
\section{SDMEs}
\begin{figure}
\begin{center}
\includegraphics[width=0.3\tw]{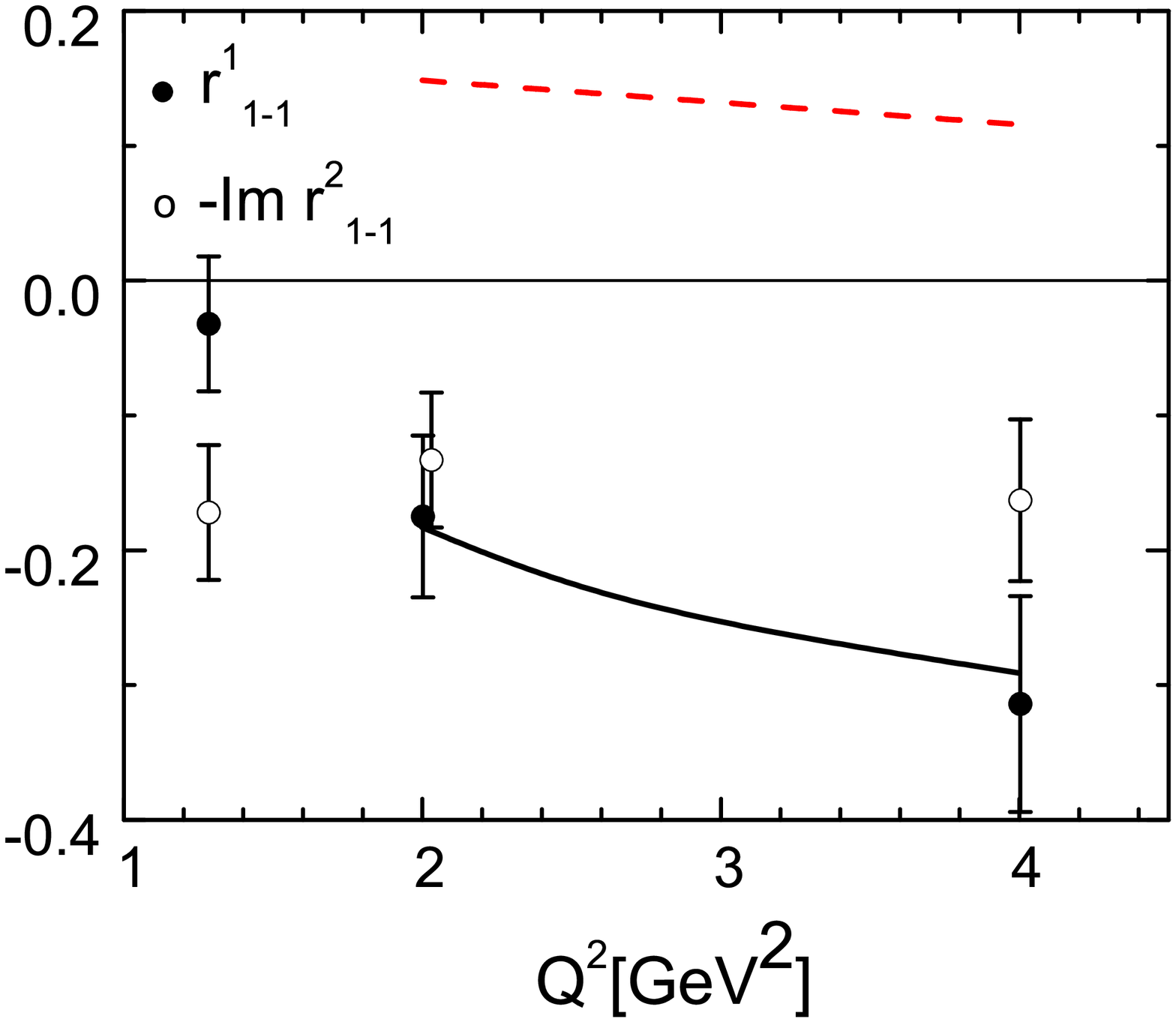}\hspace*{0.03\tw}
\includegraphics[width=0.3\tw]{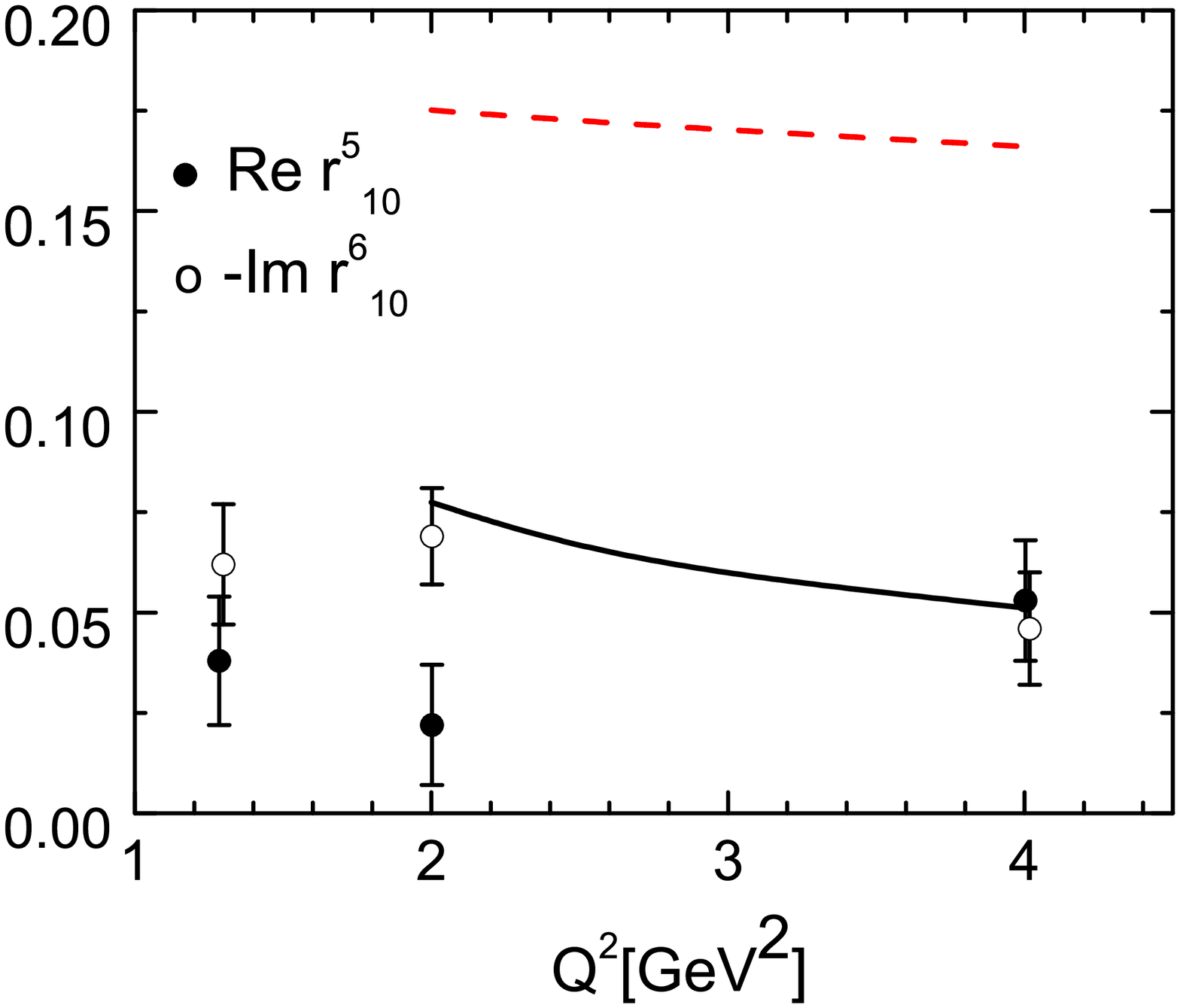}\hspace*{0.03\tw}
\includegraphics[width=0.3\tw]{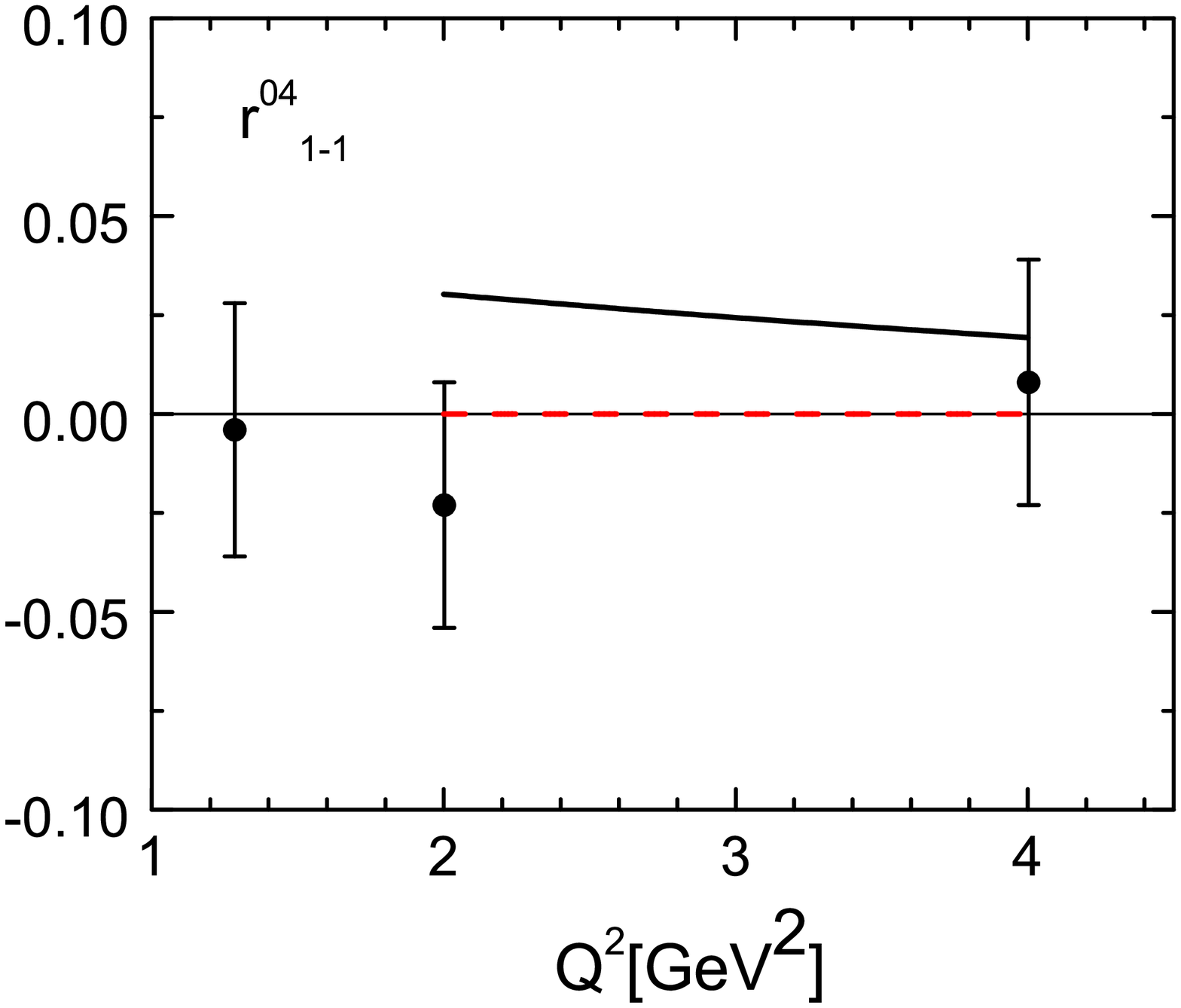}
\end{center}
{~}
\begin{center}
\includegraphics[width=0.3\tw]{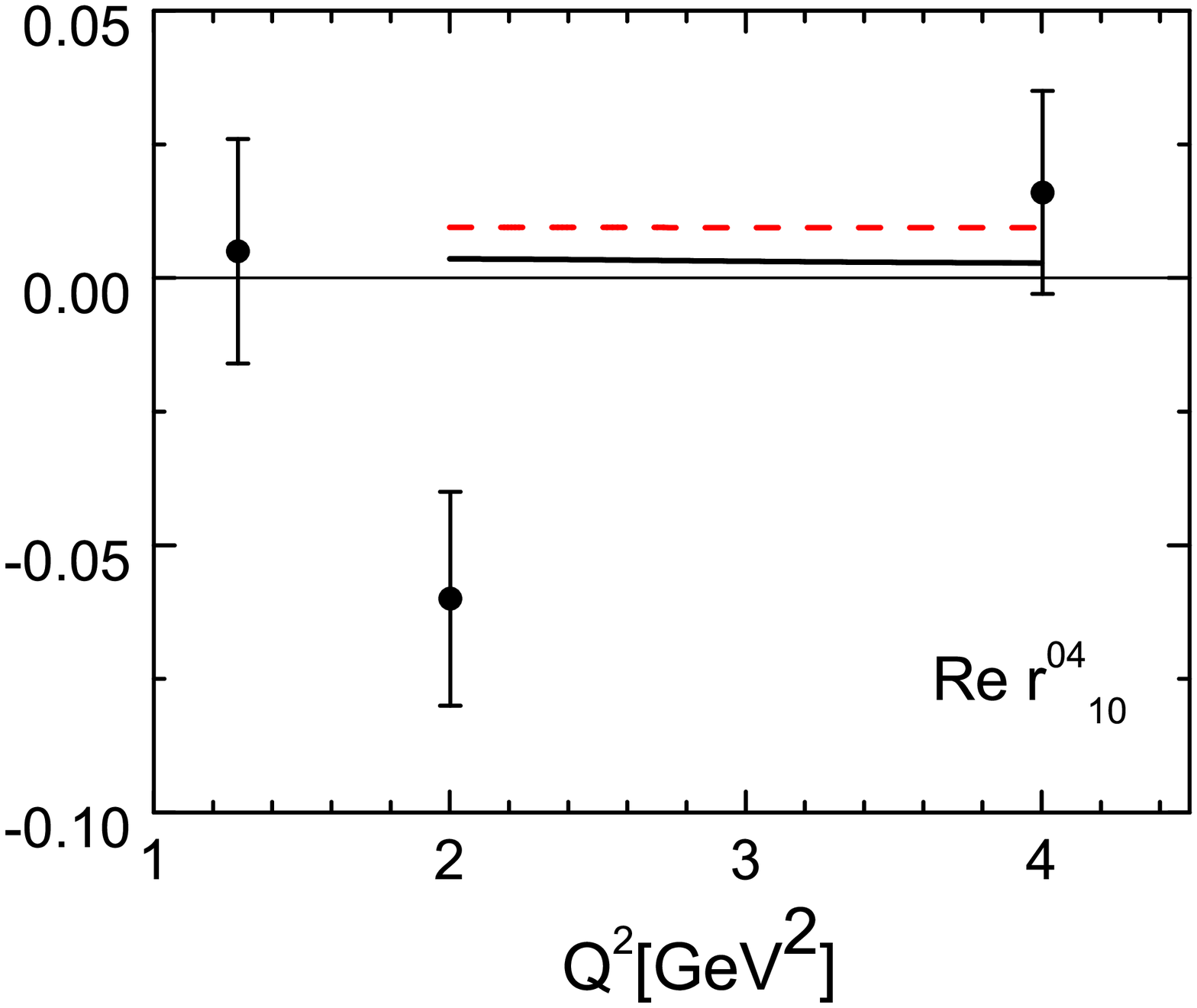}\hspace*{0.03\tw}
\includegraphics[width=0.3\tw]{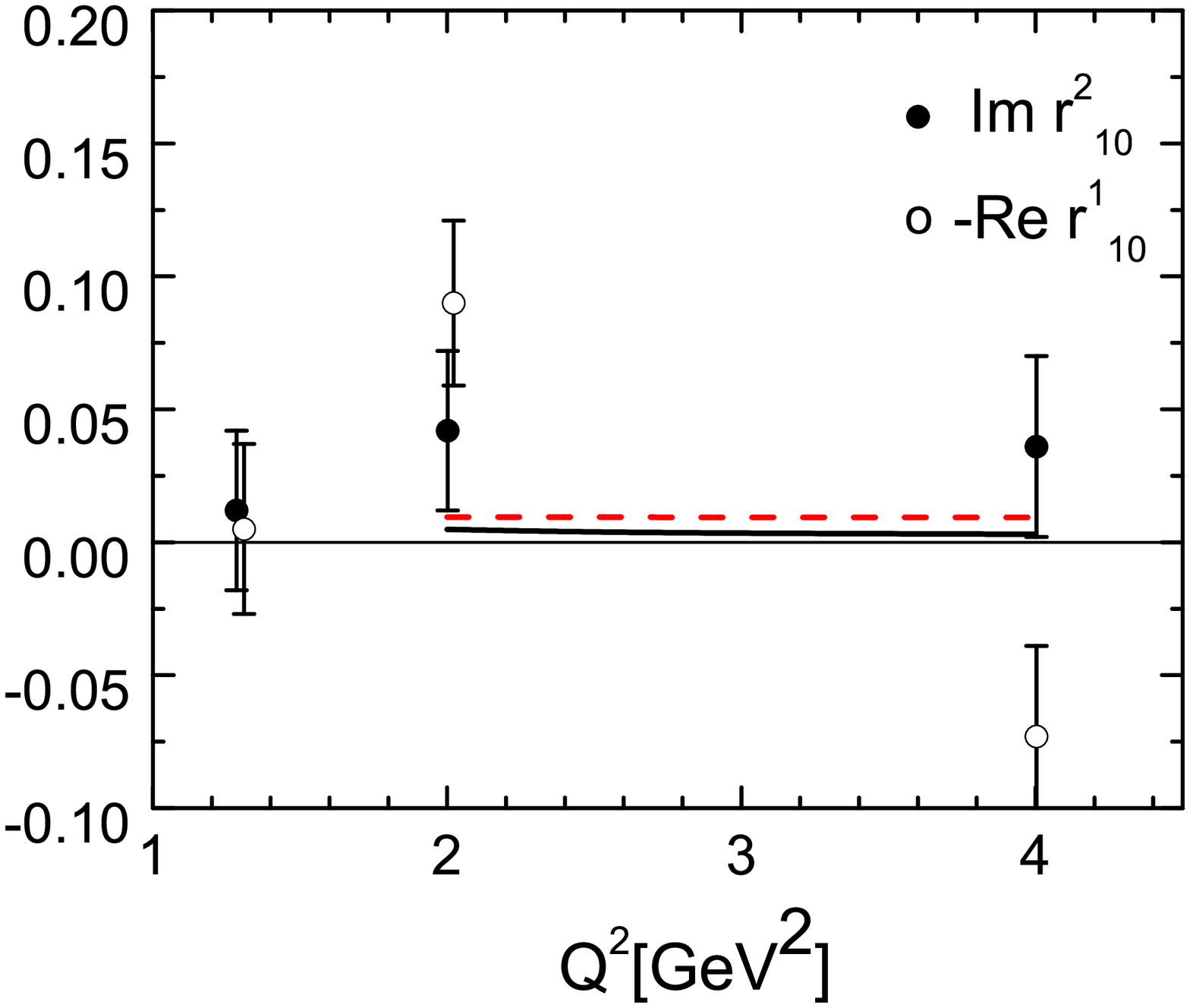}\hspace*{0.03\tw}
\includegraphics[width=0.3\tw]{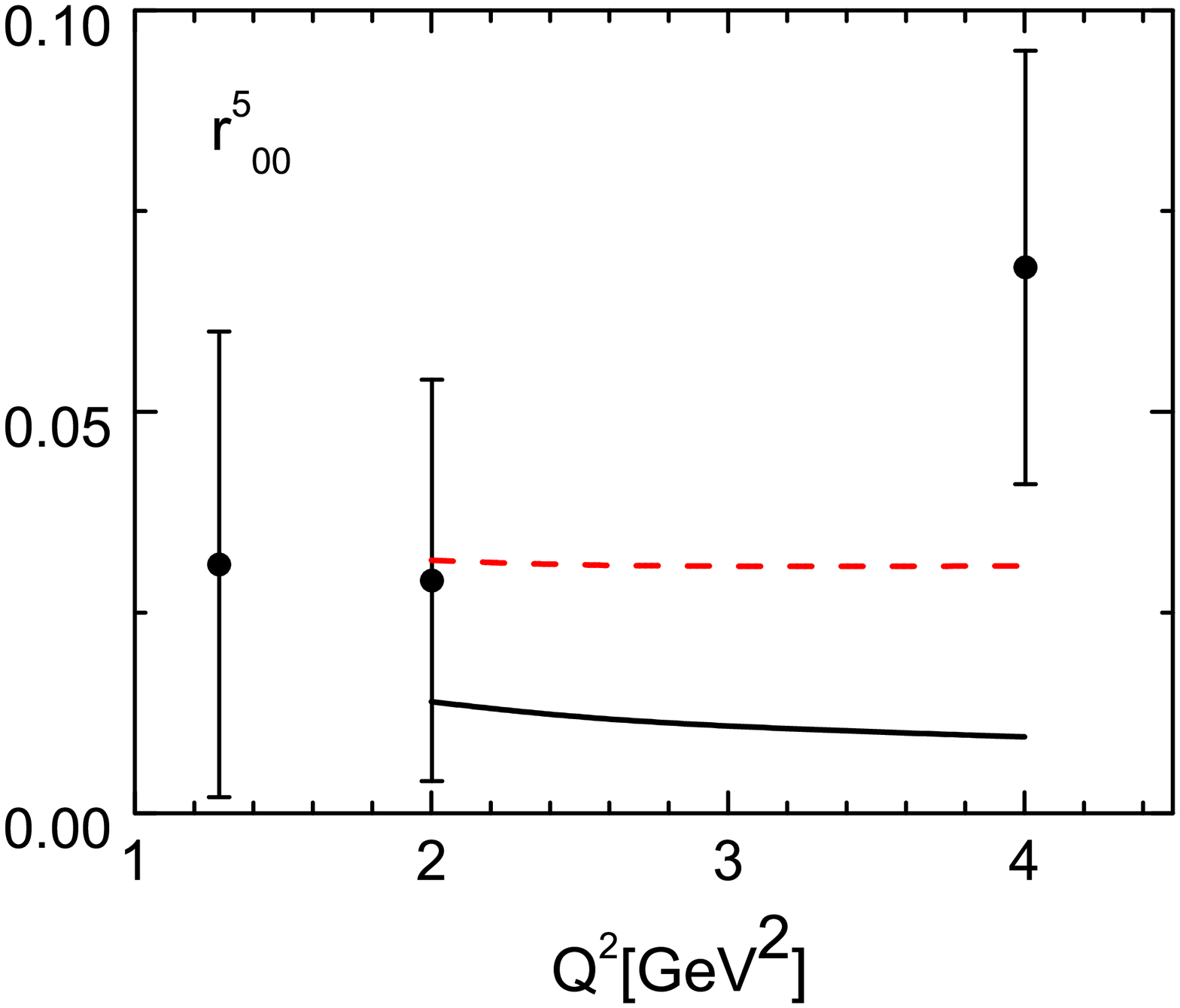}
\end{center}
{~}
\begin{center}
\includegraphics[width=0.3\tw]{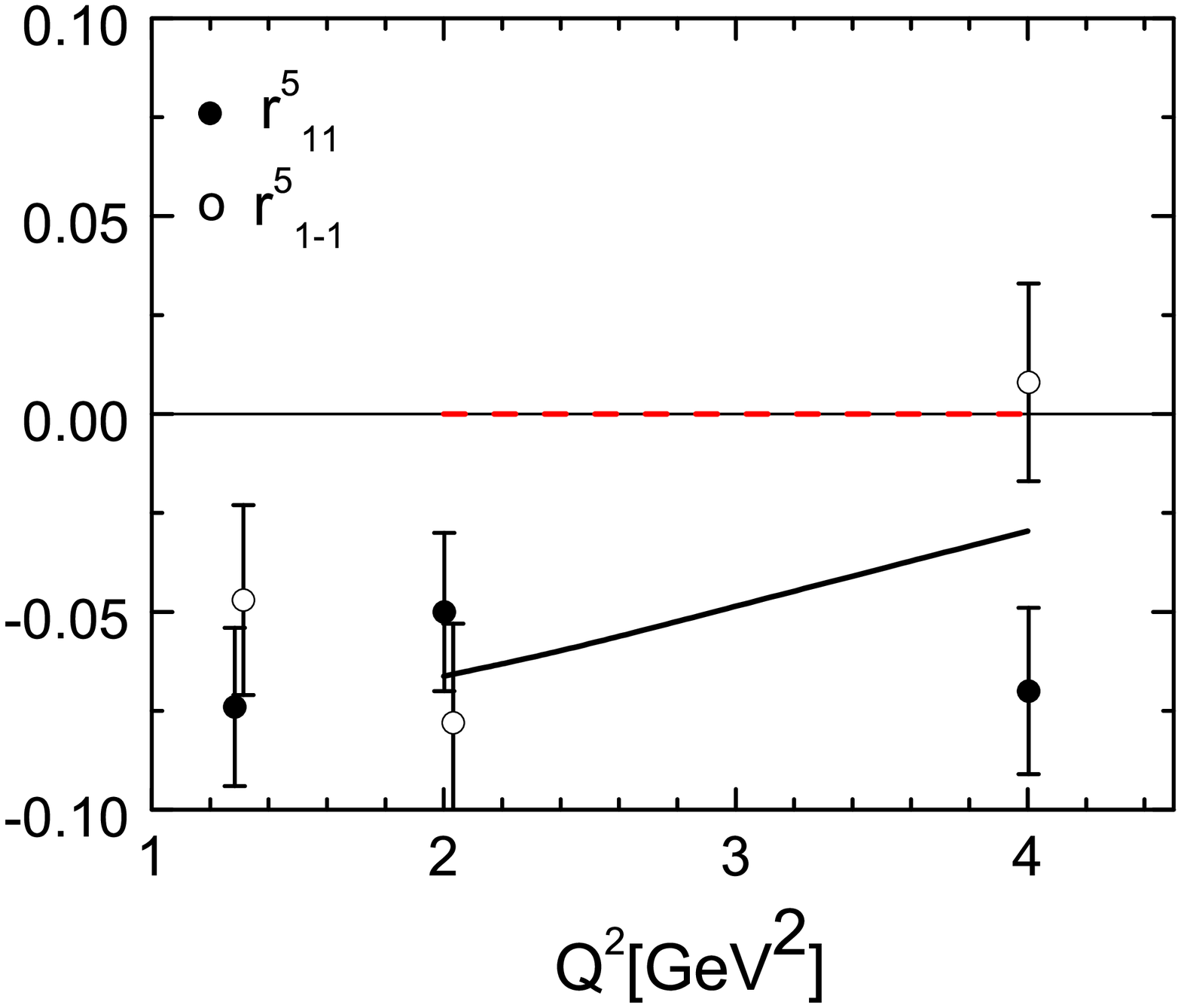}\hspace*{0.03\tw}
\includegraphics[width=0.3\tw]{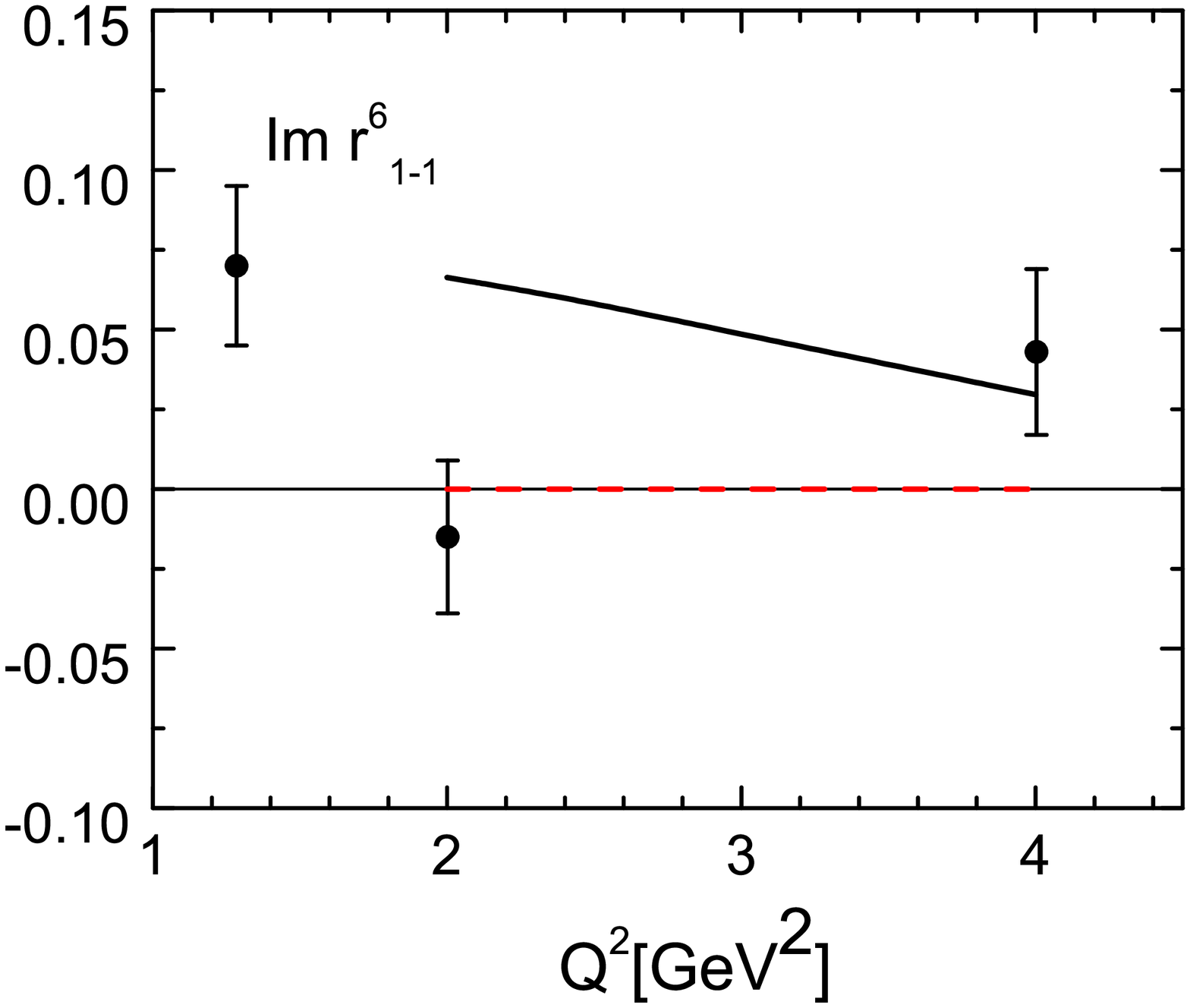}\hspace*{0.03\tw}
\includegraphics[width=0.3\tw]{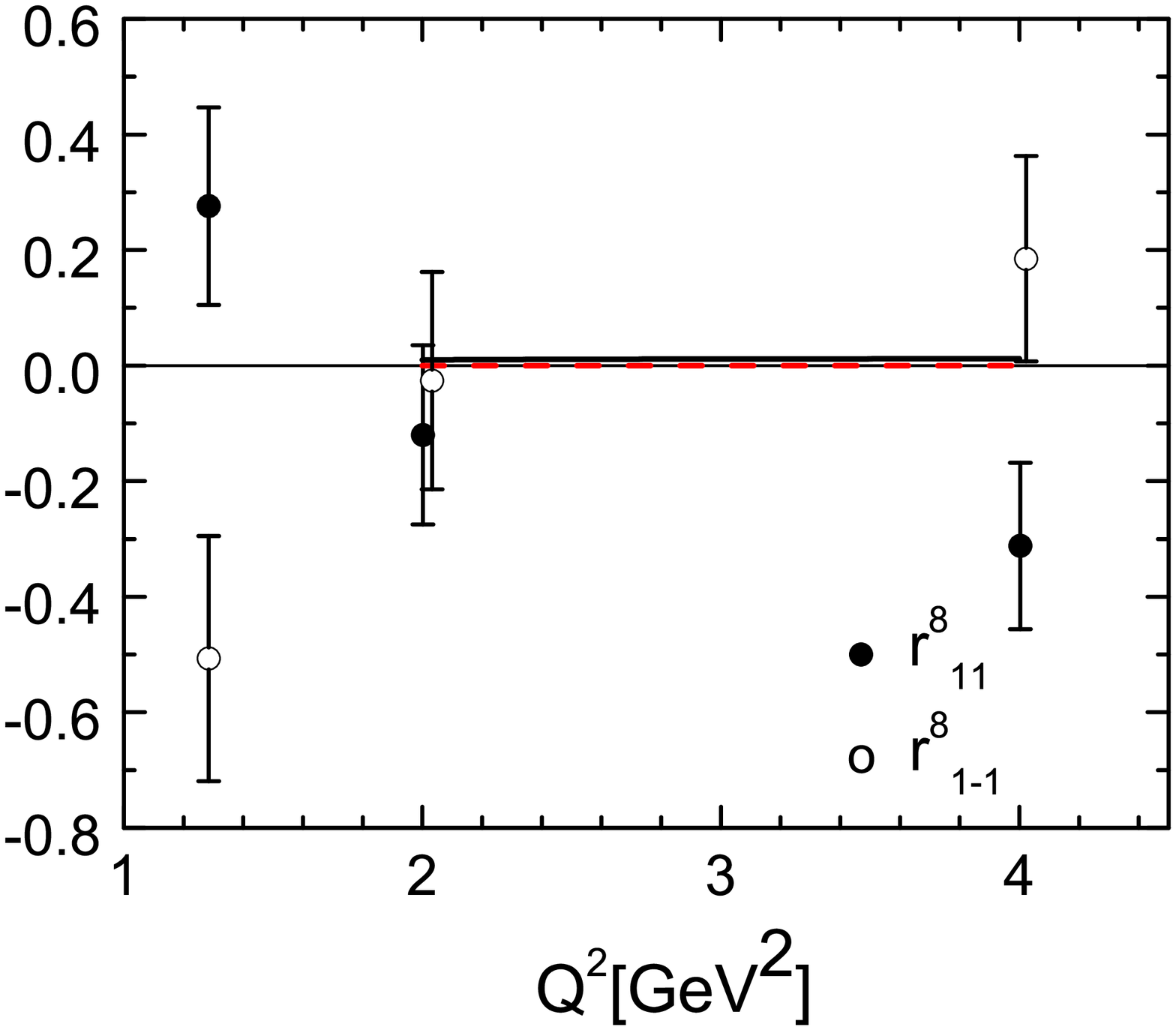}
\end{center}
\caption{Various SDMEs versus $Q^2$ at $W=4.8\,\gev$ and $t'=-0.08\,\gev^2$. Data are taken
from \ci{hermes-omega}. For other notations it is referred to Fig.\ \ref{fig:U1}.}.
\label{fig:sdme-Q}
\end{figure}

\begin{figure}
\begin{center}
\includegraphics[width=0.3\tw]{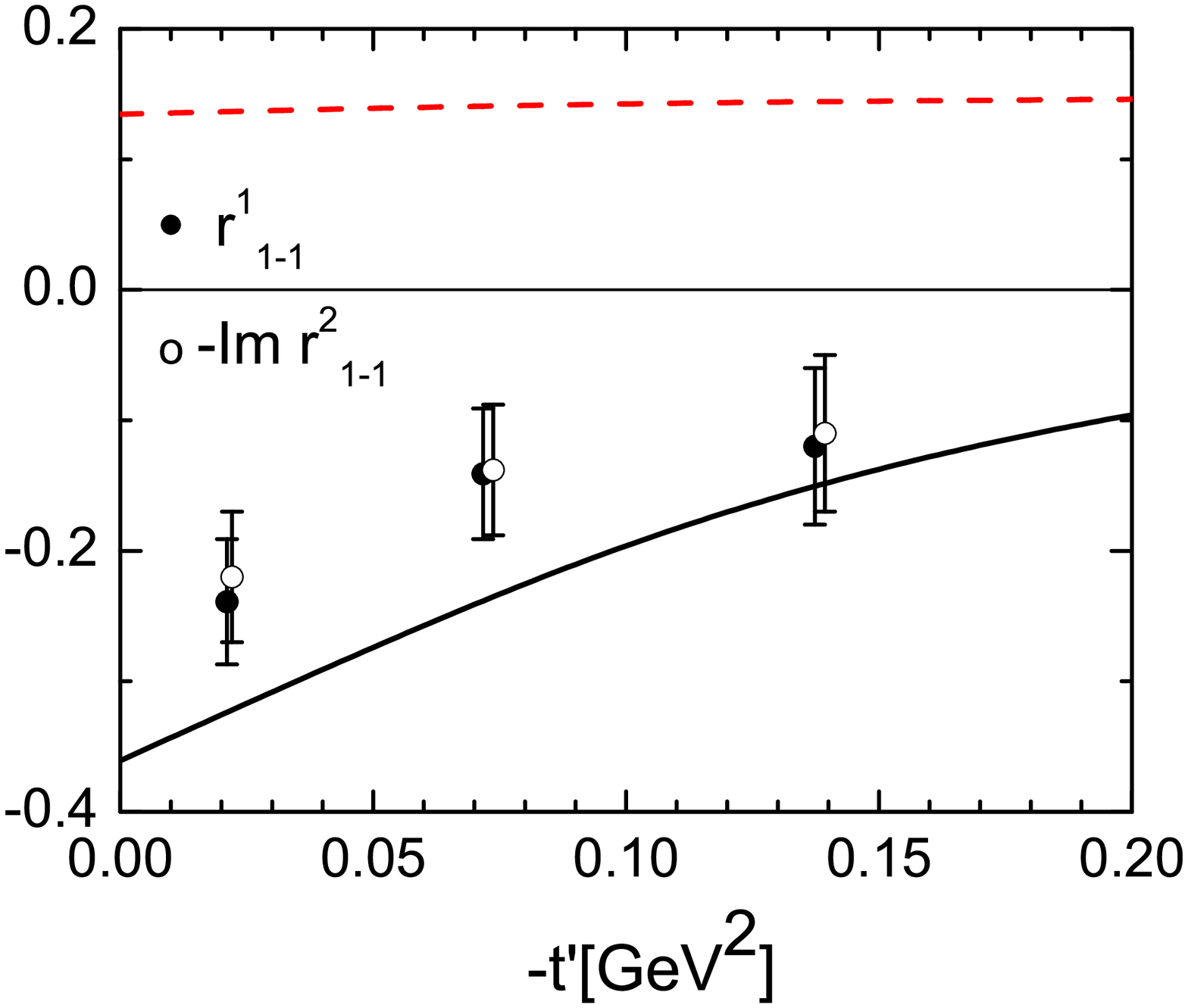}\hspace*{0.03\tw}
\includegraphics[width=0.3\tw]{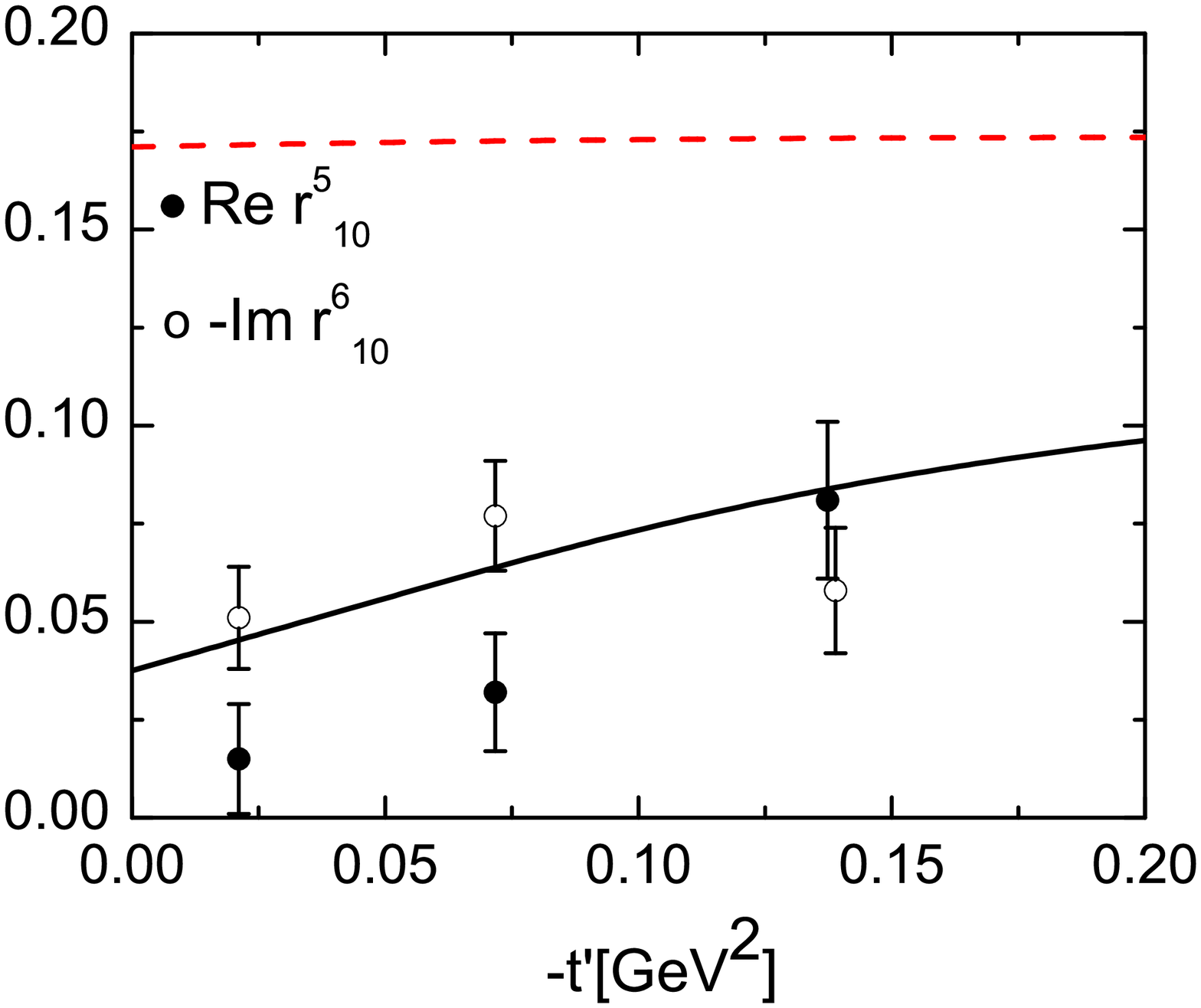}\hspace*{0.03\tw}
\includegraphics[width=0.3\tw]{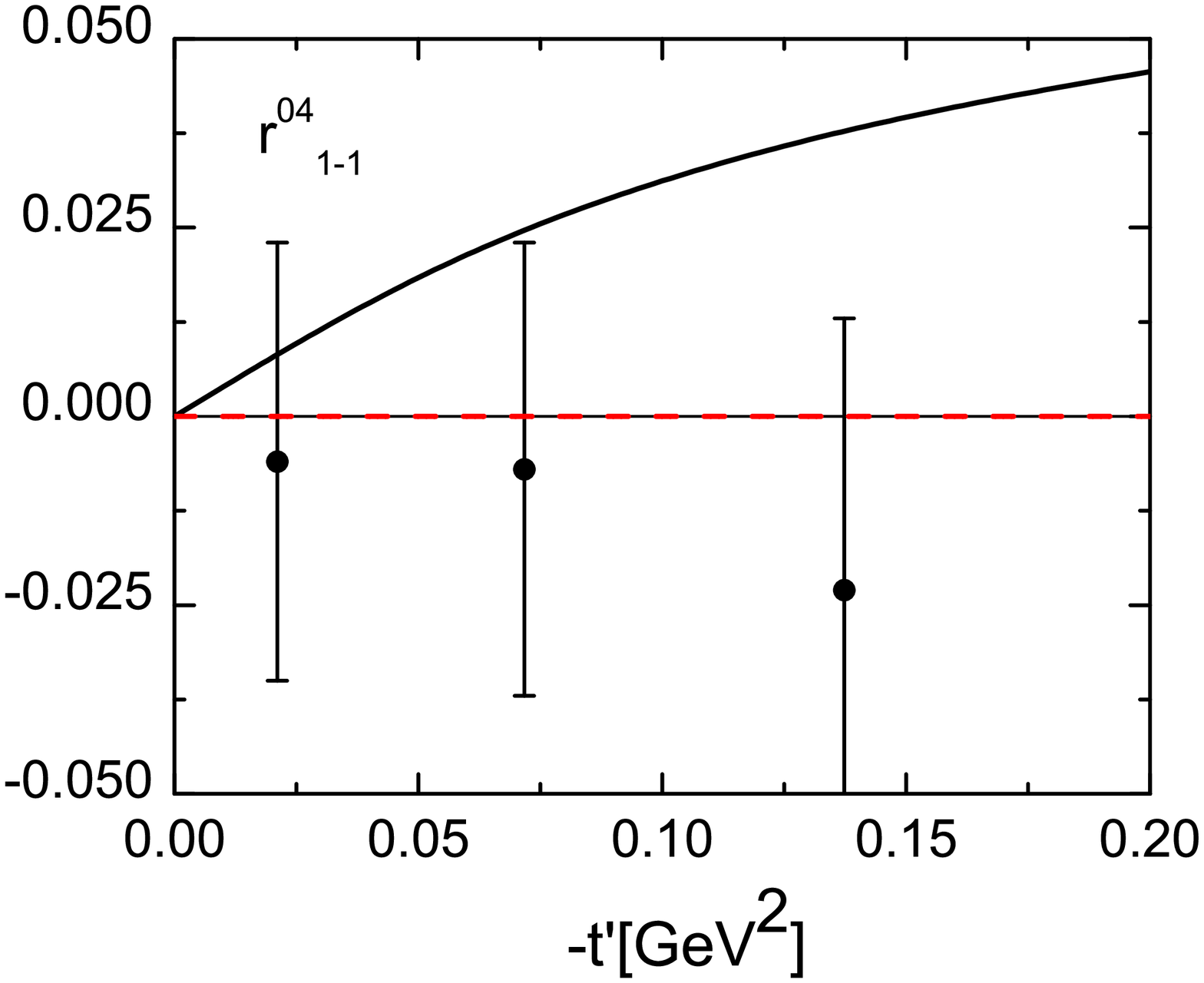}
\end{center}
{~}
\begin{center}
\includegraphics[width=0.3\tw]{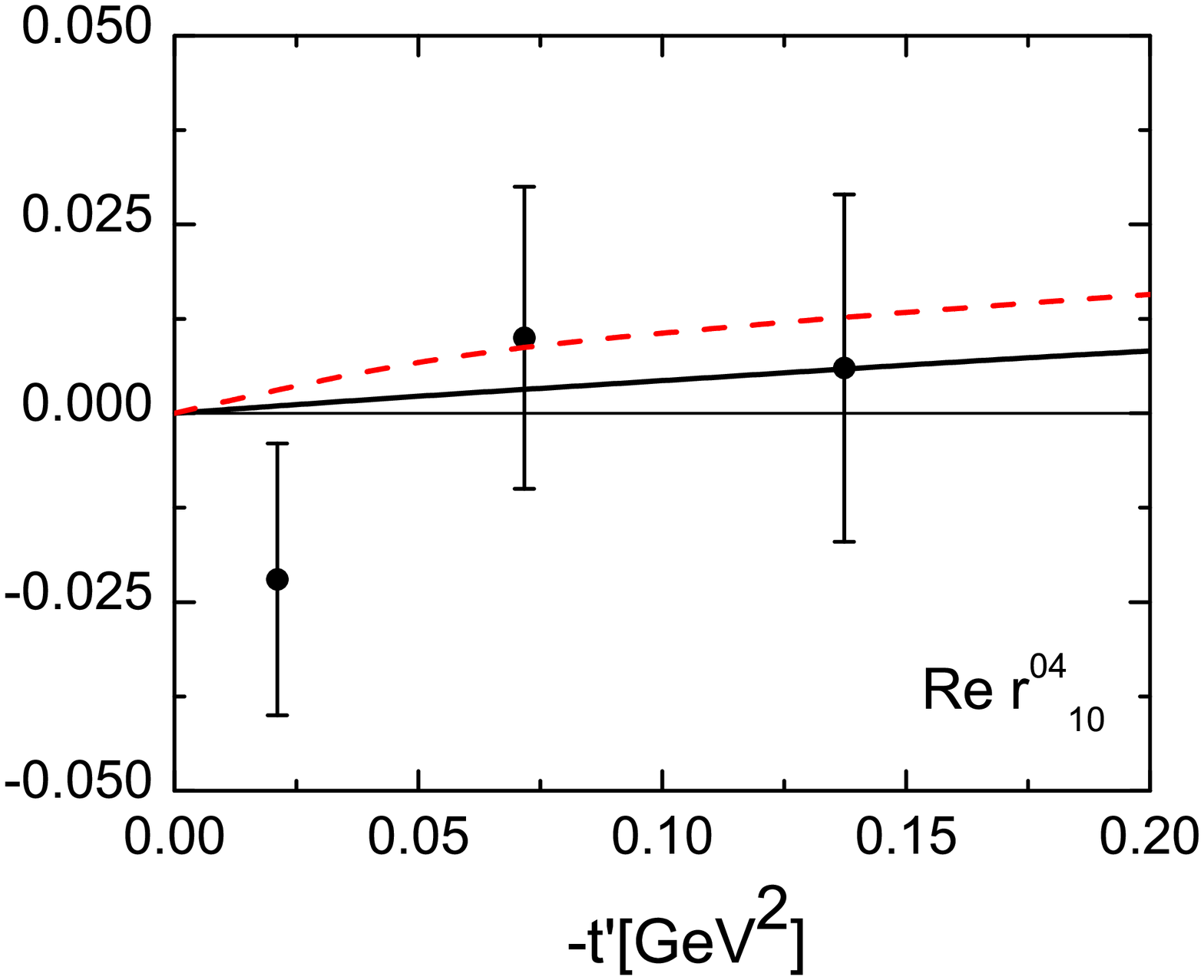}\hspace*{0.03\tw}
\includegraphics[width=0.3\tw]{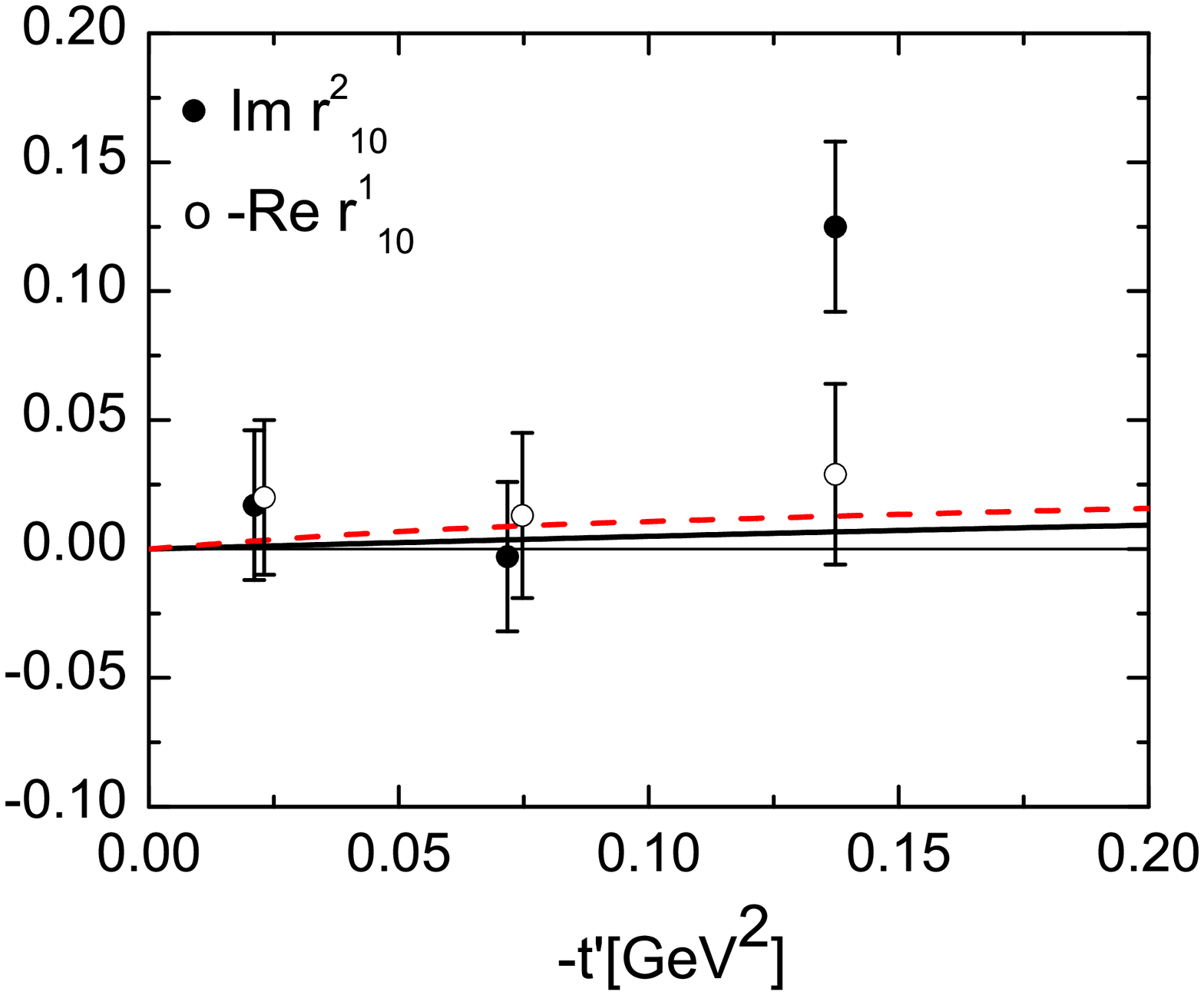}\hspace*{0.03\tw}
\includegraphics[width=0.3\tw]{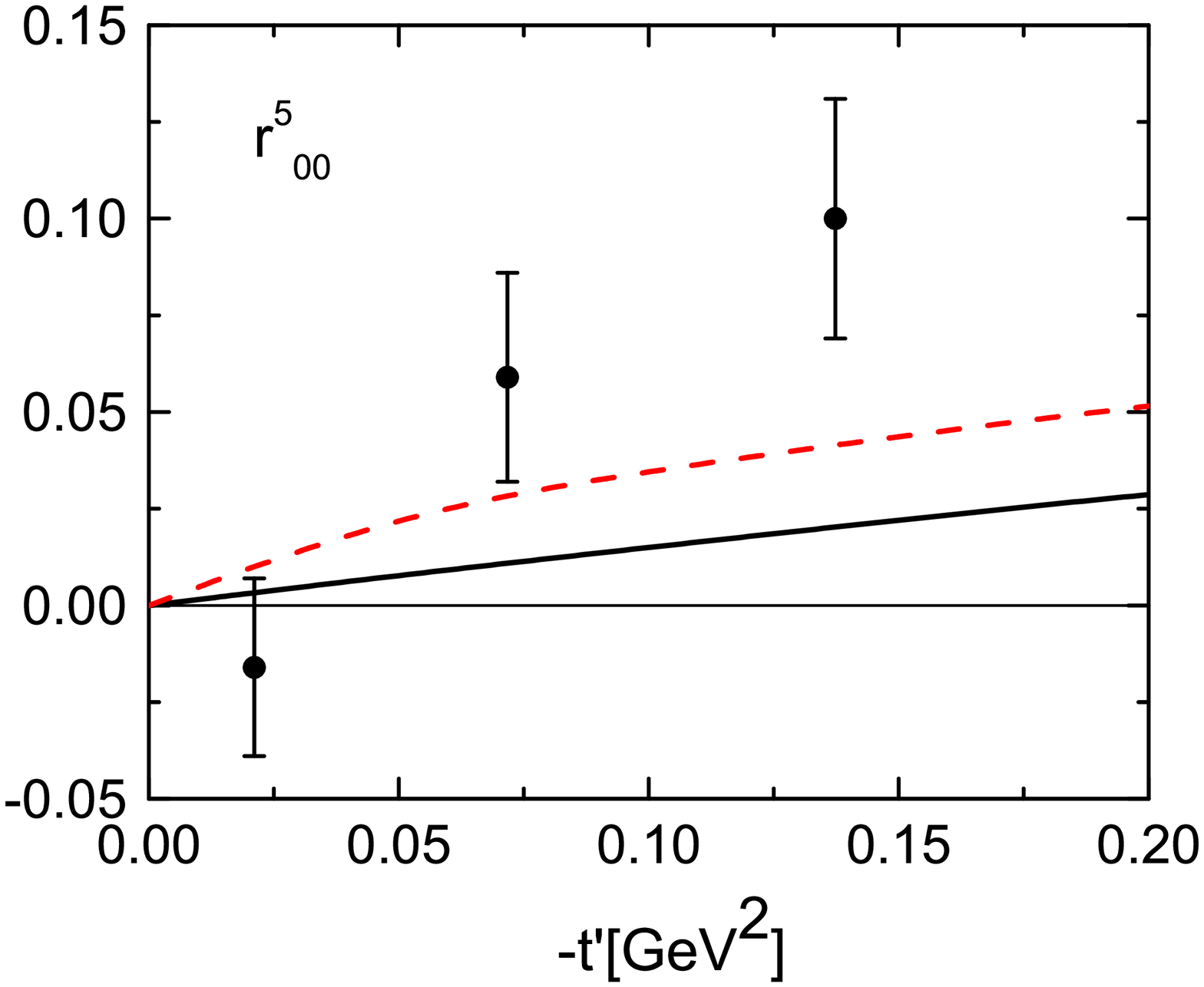}
\end{center}
{~}
\begin{center}
\includegraphics[width=0.3\tw]{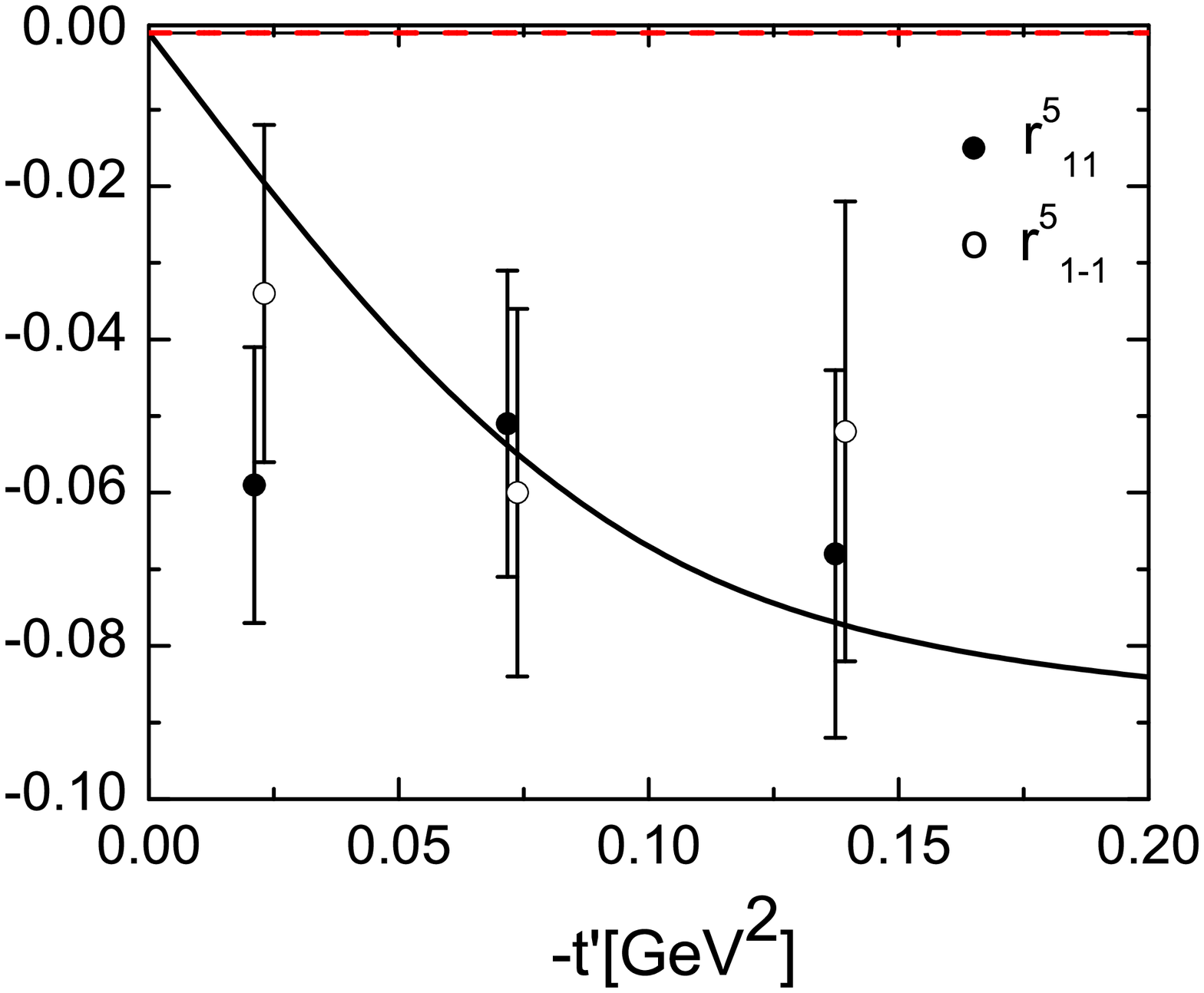}\hspace*{0.03\tw}
\includegraphics[width=0.3\tw]{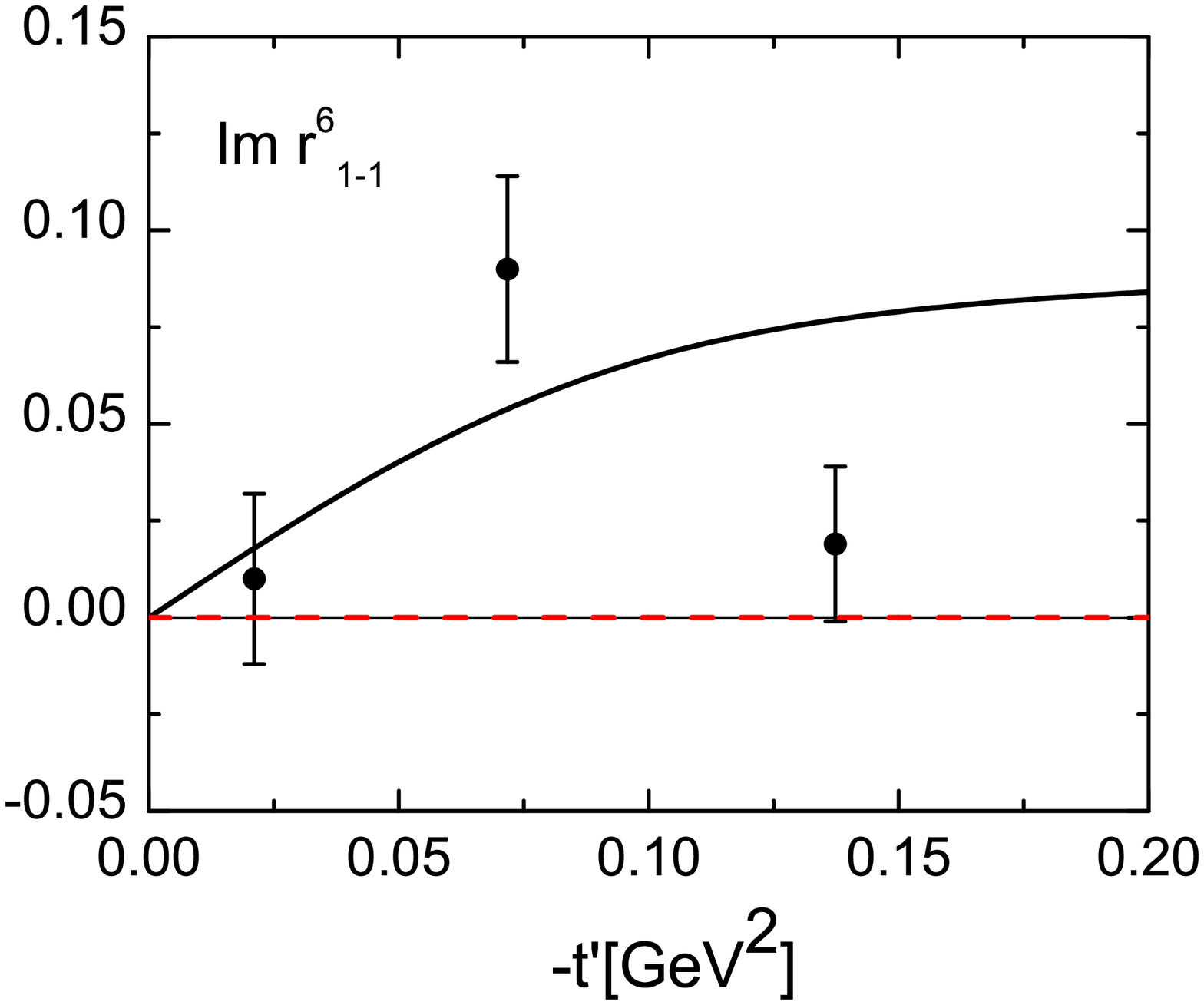}\hspace*{0.03\tw}
\includegraphics[width=0.3\tw]{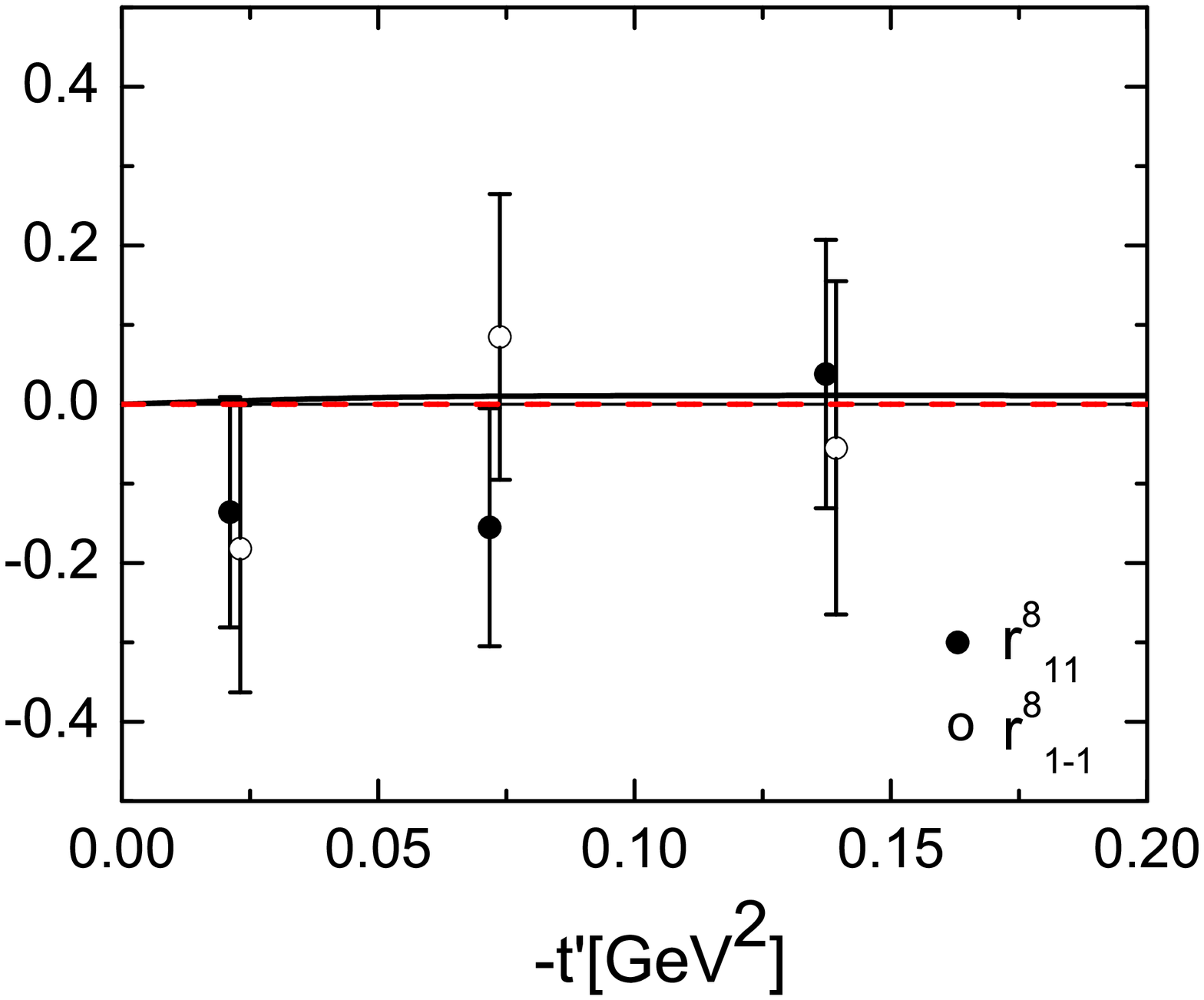}
\end{center}
\caption{Various SDMEs versus $t'$ at $W=4.8\,\gev$ and $Q^2=2.42\,\gev^2$. Data are taken
from \ci{hermes-omega}. For other notations it is referred to Fig.\ \ref{fig:U1}.}.
\label{fig:sdme-t}
\end{figure}

In this section we compare our results for the $\omega$ SDMEs with the HERMES data
\ci{hermes-omega}. Since we neglect the $\gamma^*_T\to V_{-T}^{\phantom{*}}$ transitions 
some of the SDMEs fall together, e.g.\ $r^1_{1-1}=-{\rm Im}\, r^2_{1-1}$; others 
are approximately equal. In each such case we combine the SDMEs in one plot. A number 
of SDMEs are very small or even zero. If this agrees with experiment we do not display 
these SDMEs, e.g.\ $r_{00}^1$ or ${\rm Im}\, r_{10}^3$. The remaining SDMEs are shown  
versus $Q^2$ in Fig.\ \ref{fig:sdme-Q} and versus $t'$ in Fig.\ \ref{fig:sdme-t}. In 
general we observe fair agreement between our results and the HERMES data. The 
importance of the pion pole is clearly visible, some of the SDMEs drastically change 
their values if the pion pole is neglected, for instance $r^1_{1-1}$. We stress that 
the results on the SDMEs shown in the Figs.\ \ref{fig:sdme-Q} and \ref{fig:sdme-t} are 
evaluated from the $\pi\omega$ transition form factor \req{eq:FFparametrization}, 
assuming a positive sign for it. Choosing it to be negative leads to results
which agree with the other ones within the experimental errors. 

Since  
\be
r_{00}^{04}\=\frac{d\sigma(\gamma^*_T\to V_L^{\phantom{*}})
              +\varepsilon d\sigma^N(\gamma^*_L\to V_L^{\phantom{*}})}{d\sigma}
\ee
this SDME is sufficiently well probed by the combinations of SDMEs discussed
in Sect.\ \ref{sec:partial} we do not display it here. The equality of $r^1_{1-1}$ and 
$-{\rm Im}\, r^2_{1-1}$ 
\be
r^1_{1-1}\=-{\rm Im}\, r^2_{1-1}\=\frac{d\sigma^N(\gamma^*_T\to V_T^{\phantom{*}})
                -d\sigma^U(\gamma^*_T\to V_T^{\phantom{*}})}{2d\sigma}
\ee
is in fair agreement with experiment as can be seen from the first rows of Figs.\ 
\ref{fig:sdme-Q} and \ref{fig:sdme-t}. The (class B) SDMEs ${\rm Re}\, r^5_{10}$ and 
${\rm Im}\, r^6_{10}$ fall practically together since they are dominated by the real part of
\be 
\sum_{\nu'}{\cal M}^N_{+\nu',++}{\cal M}^{N*}_{0\nu',0+}\,.
\ee
The imaginary part of this interference term controls 
${\rm Im}\, r^7_{10}\simeq {\rm Re}\, r^8_{10}$. It is very small in our approach in agreement 
with experiment within admittedly large errors. A little difference between 
${\rm Re}\, r^5_{10}$ and ${\rm Im}\, r^6_{10}$ is caused by the term 
${\rm Re}\, \big[{\cal M}^U_{+-,0+}{\cal M}^*_{0-,++}\big]$ appearing in ${\rm Im}\, r^6_{10}$. 
This term is proportional to the $\pi\omega$ transition form factor and therefore depends on 
its sign. Given the experimental errors this has however no noticeable consequences since 
this term is so small. Thus, SDMEs of class B depend on the pion pole only through the 
normalization $d\sigma/dt$. The SDME $r^{04}_{1-1}$ measures the pion-exchange cross section 
$d\sigma^{\rm pole}(\gamma_L^*\to V_T^{\phantom{*}})$ (see \req{eq:pole-cross-section}) 
\be
r^{04}_{1-1}\=\frac{\varepsilon}{2}\,\frac{d\sigma^U(\gamma^*_L\to V_T^{\phantom{*}})}{d\sigma}
\ee
Also for this small SDME which asymptotically decreases as $Q^{-4}$, we find reasonable,
although not perfect agreement with the data. 

The (class C) SDMEs ${\rm Re}\, r^{04}_{10}$, ${\rm Re}\, r^1_{10}=-{\rm Im}\, r^2_{10}$ and 
$r^5_{00}$, shown in the second rows of Figs.\ \ref{fig:sdme-Q} and \ref{fig:sdme-t}, are 
sensitive to the transversity GPDs $H_T$ and $\bar{E}_T$ feeding the 
$\gamma^*_T\to V_L^{\phantom{*}}$ amplitudes \ci{GK7}. For these SDMEs the pion pole mainly 
affects the normalization $d\sigma/dt$. The SDME $r^5_{00}$ is dominated by  
${\rm Re}\,{\cal M}^N_{0+,0+} {\cal M}^{N*}_{0+,++}$, the other three by 
${\rm Re}\,{\cal M}^N_{++,++}{\cal M}^{N*}_{0+,++}$. All these SDMEs

\noindent  are rather small and 
in reasonable with the HERMES data which are subject to rather large errors. An exception
is the $t$-dependence of $r^5_{00}$. The good agreement of this SDME for $\rho^0$ 
production with our results \ci{GK7} makes it difficult to improve the results for 
$r^5_{00}(\omega)$. 

In the last rows of Figs.\ \ref{fig:sdme-Q} and \ref{fig:sdme-t} the (class D)
SDMEs $r^5_{11}=r^5_{1-1}=-{\rm Im}\, r^6_{1-1}$ and $r^8_{11}=r^8_{1-1}={\rm Im}\, r^7_{1-1}$
are displayed which measure the real and imaginary part of
\be 
\sum_{\nu'}{\cal M}^U_{+\nu',++}{\cal M}^{U*}_{+\nu',0+}\,,
\label{eq:interTLU}
\ee
respectively. Up to a small contribution from $\widetilde H$ to ${\cal M}^U_{++,++}$ 
these SDMEs probe the pion-exchange amplitudes \req{eq:hel-ampl}. Obviously, these SDMEs 
are zero if the pion-pole contribution is neglected. The imaginary part of 
\req{eq:interTLU} is just the interference term of the $\widetilde H$ and the 
pion-pole contributions. It is non-zero although very small. Moreover, it is 
$\sim g_{\pi\omega}$ and, hence, changes sign together with the form factor. Inspection 
of Figs.\ \ref{fig:sdme-Q} and \ref{fig:sdme-t} reveals that the HERMES data \ci{hermes-omega} 
do not fix the sign of the transition form factor.

We refrain from showing predictions for the $\omega$ SDMEs at $W=3.5$ and $8\,\gev$
typical for the upgraded JLab and the COMPASS experiment, respectively. The results 
at these energies look similar to those at $W=4.8\,\gev$. At $3.5\,\gev$ the results 
are further away from those obtained under neglect of the pion-pole contribution, at 
$8\,\gev$ they are closer.

\section{Spin asymmetries}
In \ci{GK7} we have investigated various spin asymmetries and it is now obligatory
to check whether the results presented in \ci{GK7} will be substantially changed 
by the inclusion of the pion pole or not. In this connection we can also examine
whether there are asymmetries which are sensitive to the sign of the $\pi V$
transition form factor. Expressing the asymmetries for longitudinal and transverse
beam and target polarizations, $A_{UT}, A_{LT}, A_{LU}, A_{UL}, A_{LL}$, in terms of 
helicity amplitudes \ci{GK7,sapeta}, we find two potentially large interference
terms with the pion-pole contribution 
\be
  {\cal M}^{N*}(\gamma^*_T\to V^{\phantom{*}}_T)
              {\cal M}^U(\gamma^*_{L,T}\to V^{\phantom{*}}_T)
\label{eq:inter1}
\ee
and 
\be
{\cal M}^{U*}(\gamma^*_T\to V^{\phantom{*}}_T)
                      {\cal M}^U(\gamma^*_{L,T}\to V^{\phantom{*}}_T)
\ee
The imaginary part of the latter interference term reduces to that of the contributions 
from $\widetilde H$ and the pion pole. This term as well as the one given in \req{eq:inter1} 
change sign with the transition form factor and mainly affect $A_{UT}$ and $A_{UL}$. The pion 
pole affects all spin asymmetries through the normalization, the unseparated cross 
section. This effect is however substantial only for $\omega$ production at energies less 
than about $6\,\gev$. Note that the term 
\be
           {\rm Re} \sum_{\nu'}{\cal M}^{U*}_{+-\nu',++}{\cal M}^U_{+\nu',0+}
\ee
contributing to the $\cos{\phi_s}$ modulation of $A_{LT}$ is zero~\footnote{
$\phi_s$ is the orientation of the target spin vector with respect to the lepton plane and
$\phi$ specifies the azimuthal angle between the lepton and the hadron plane.}.

\begin{figure}
\begin{center}
\includegraphics[width=0.45\tw]{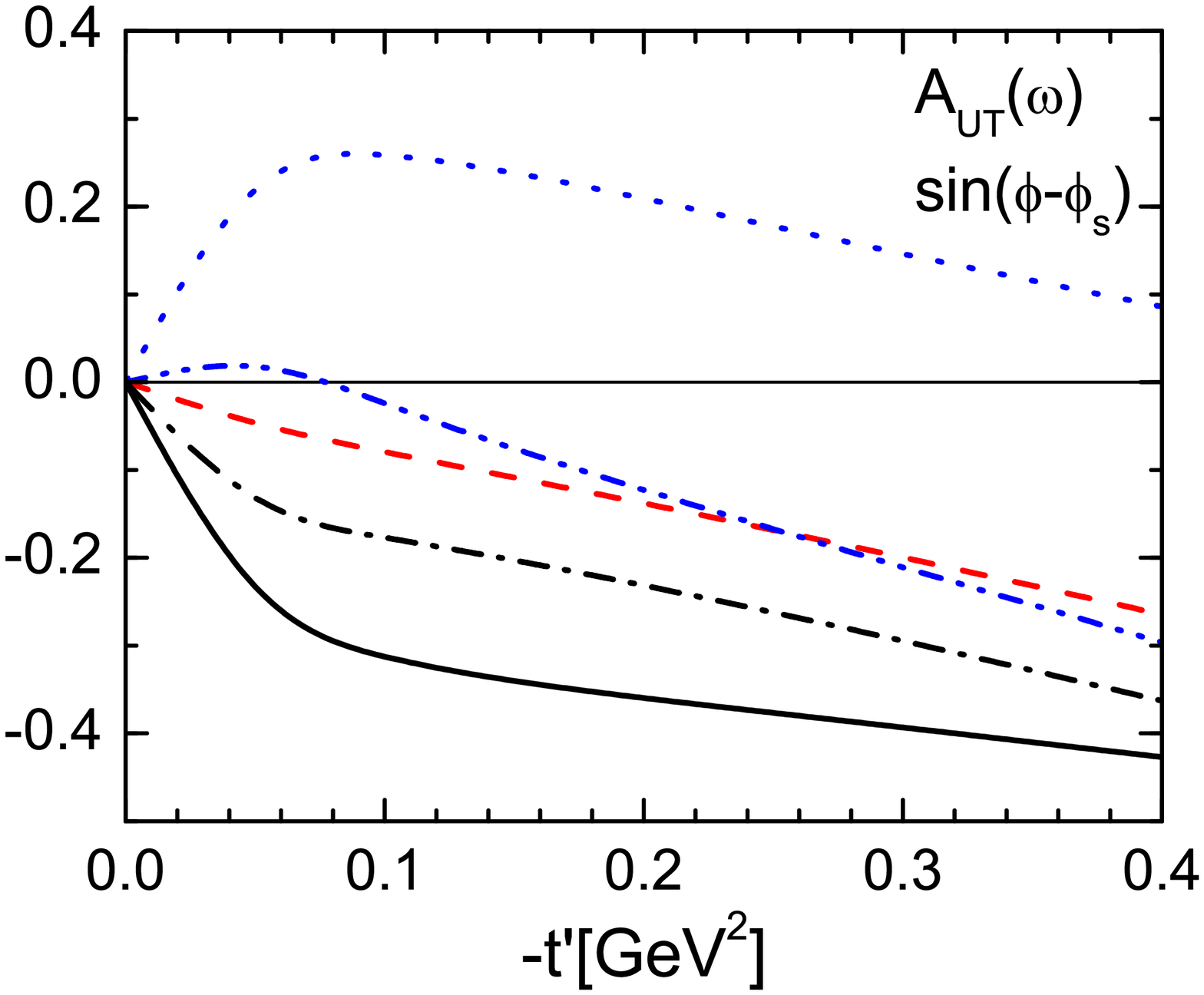}\hspace*{0.05\tw}
\includegraphics[width=0.45\tw]{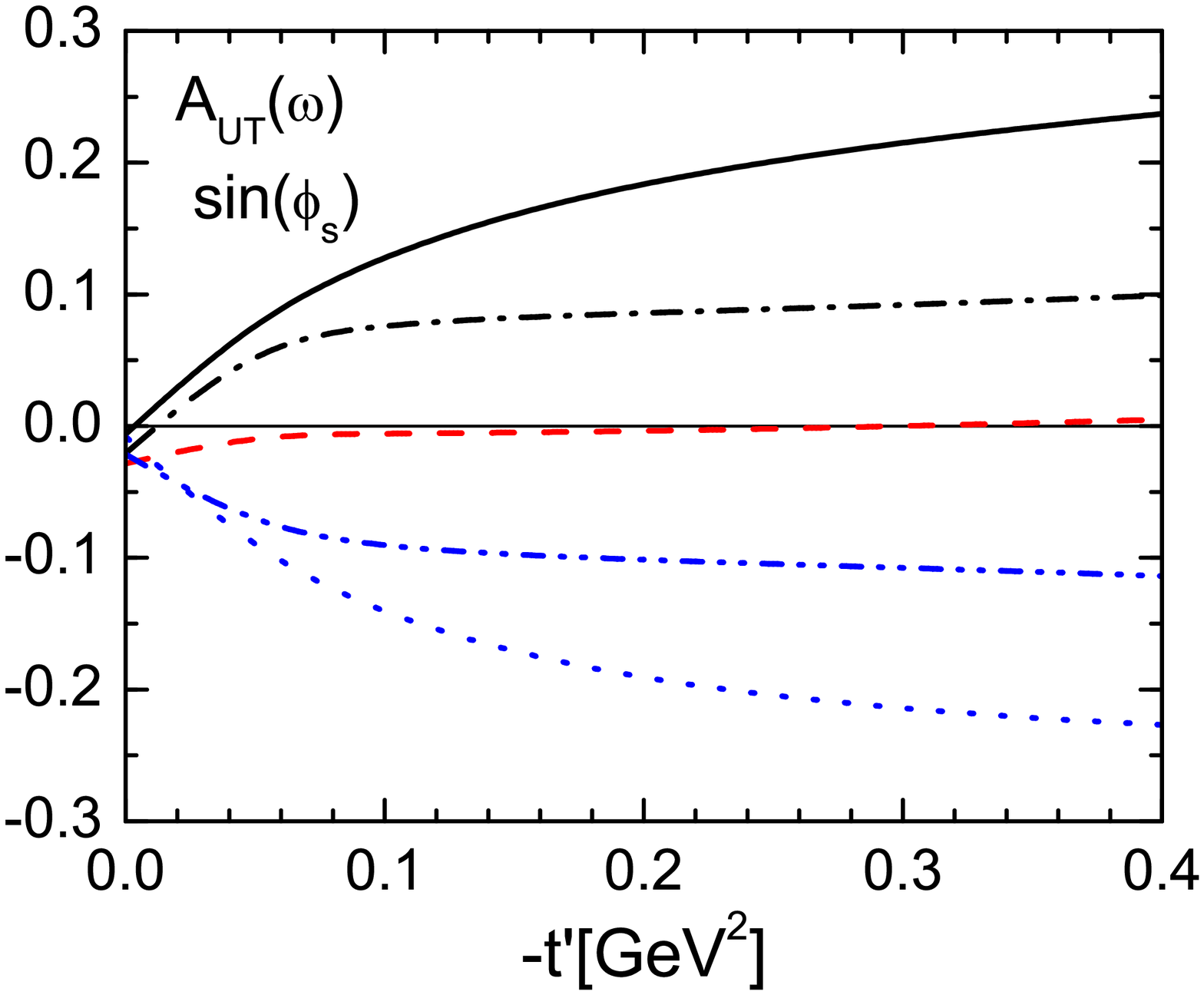}
\end{center}
\caption{$\sin(\phi-\phi_s)$ (left) and $\sin \phi_s$ (right) modulations of $A_{UT}$
versus $t'$ for $\omega$ production at $W=4.8\,\gev$ and $Q^2=2.42\,\gev^2$. 
The solid (dotted) lines represent our results from the handbag approach with a 
positive (negative) $\pi\omega$ form factor; the dashed lines are the results without 
the pion pole. The dash-dotted (dash-dot-dotted) lines are predictions at $W=8\,\gev$
and $Q^2=2.42\,\gev^2$ with a postive (negative) $\pi\omega$ form factor.
The $\varepsilon$-dependent prefactors of the asymmetries are included, see \ci{
GK7,sapeta}.}
\label{fig:asymmetries}
\end{figure}

Two examples of our predictions for asymmetries in $\omega$ leptoproduction are shown in
Fig.\ \ref{fig:asymmetries}. The effects of the pion pole are particularly large for these
asymmetries and the sign of the $\pi \gamma$ form factor matters.  

For $\rho^0$ production only little effects are generated by the pion pole. The agreement
of our previous results with the experimental data on $A_{UT}$ and $A_{LT}$ 
\ci{hermes-rho-07,compass-rho,compass-rho-12} remains true. For $\omega$
production at $W\simeq 8\,\gev$ the pion pole still affects somewhat the asymmetries,
in particular the $\sin(\phi-\phi_s)$ and $\sin \phi_s$ modulations of the transverse
target asymmetry $A_{UT}$ which are even sensitive to the sign of the $\pi\omega$ transition 
form factor (see Fig.\ \ref{fig:asymmetries}).  

\section{Summary}
In the present work we have analyzed the data on the SDMEs of the omega meson 
measured by the HERMES collaboration \ci{hermes-omega} recently. In this analysis
we have made use of the handbag approach and exploited a set of GPDs extracted by us
from data on leptoproduction of $\rho^0, \phi$ and $\pi^+$ mesons \ci{GK3,GK5,GK6}. 
In addition we have allowed for the pion pole which, as it turns out, plays a very
important role in $\omega$ production. The coupling of the exchanged pion to the
proton is known from other sources (see, for instance, \ci{GK5,GK6}) while that
to the virtual photon and the $\omega$ meson, i.e.\ the $\pi\omega$ transition form
factor, is fixed from the $\omega$ SDMEs. With the exception of this form factor
there is no free parameter in our analysis. We have obtained reasonable values for
this form factor and in general a fair description of the HERMES data on $\omega$
production for $Q^2\gsim 2\,\gev^2$. For $Q^2$ less than 
$2\,\gev^2$ the GPDs and the handbag approach are not probed against experiment.
There are various approximations made in the handbag approach which become inaccurate
at low $Q^2$, e.g.\ neglect of contributions of order $t/Q^2$, target mass corrections
or higher-order perturbative corrections. As shown in \ci{diehl-kugler} the NLO 
corrections to the handbag amplitudes become large at low skewness and low $Q^2$
in the collinear limit. In this situation a resummation is required which seem to 
reduce the perturbative corrections drastically \ci{ivanov,kirchner}. Implications
of these theoretical findings for the modified perturbative approach we are using
in the calculation of the partonic subprocesses, in which quark transverse momenta are
retained and Sudakov effects in next-to-leading log approximations are taken into account,
are unclear. Nevertheless, ignoring these problems and working out the SDMEs for
$Q^2<2\,\gev^2$, we find agreement with the HERMES data on the same level of quality
as for larger $Q^2$. We stress that the SDMEs for the $\omega$ meson do not fix the sign
of the $\pi\omega$ transition form factor, some of the spin asymmetries for $\omega$
leptoproduction are however sensitive to this sign.
 
We have also commented on the role of the pion pole in $\rho^0$ and $\phi$ 
leptoproduction. In the latter case, the pion pole only contributes through $\omega - \phi$
mixing and leads to tiny effects which are negligible in practice. For $\rho^0$
production the pion pole contribution is small because of \req{eq:charge-ratio}.
It enhances the cross section by about $2\%$ and is visible only in some of the SDMEs
or combinations of SDMEs like $U_1$ and $P$.    

{\it Acknowledgements:} The authors are grateful to the HERMES collaboration for making
available to us their data on the $\omega$ SDMEs prior to publication. One of us (P.K.) 
also thanks Adrien Besse, Herv\'e Moutarde, Wolf-Dieter Nowak, Franck Sabatie and Oleg
Teryaev for helpful discussions. The work is supported in part by the Russian Foundation
of Basic Research, grant 12-02-00613, by the Heisenberg-Landau program and by the BMBF,
contract number OR 06RY9191.

\vskip 10mm

\end{document}